\documentclass[%
 reprint,
 amsmath,amssymb,
 aps,
floatfix,
]{revtex4-1}

\usepackage{graphicx}% Include figure files
\usepackage{dcolumn}% Align table columns on decimal point
\usepackage{bm}% bold math

\usepackage{qcircuit}
\usepackage[caption=false]{subfig}

\usepackage{algorithm}
\usepackage[noend]{algpseudocode}
\algrenewcommand\algorithmicdo{}

\makeatletter
\renewcommand{\ALG@name}{Procedure}
\makeatother

\newcommand{\switch}{%
  \mathcode`+=\numexpr\mathcode`+ + "1000\relax % turn + into a relation
  \mathcode`*=\numexpr\mathcode`* + "1000\relax
}

\usepackage[linktocpage=true,
  colorlinks=true, 
  pdfborder={0 0 0},
  linkcolor=blue,
  citecolor=red,
  filecolor=yellow,
  urlcolor=blue,
  bookmarks,
  pdfauthor={},
]{hyperref}

\begin{document}

\preprint{APS/123-QED}

\title{
Linear-response functions of molecules on a quantum computer:\\
Charge and spin responses and optical absorption
}

\author{Taichi Kosugi}
%\email{kosugi.taichi@gmail.com}

\author{Yu-ichiro Matsushita}
\affiliation{
Laboratory for Materials and Structures,
Institute of Innovative Research,
Tokyo Institute of Technology,
Yokohama 226-8503,
Japan
}

\date{\today}

\begin{abstract}
We propose a scheme for the construction of linear-response functions of an interacting electronic system via quantum phase estimation and statistical sampling on a quantum computer.
By using the unitary decomposition of electronic operators for avoiding the difficulty due to their nonunitarity, 
we provide the circuits equipped with ancillae for probabilistic preparation of qubit states on which the necessary nonunitary operators have acted.
We perform simulations of such construction of the charge and spin response functions and photoabsorption cross sections for C$_2$ and N$_2$ molecules by comparing with the results from full configuration-interaction calculations.
It is found that the accurate detection of subtle structures coming from the weak poles in the response functions requires a large number of measurements.
\end{abstract}

\maketitle 

\section{Introduction}

% electronic-structure calculations using quantum computers

Since the information carrier of a programmable quantum computer is a set of qubits that exploits the principle of superposition,
essentially parallel algorithms can exist and perform computation for classically formidable problems.\cite{Nielsen_and_Chuang, bib:4935}
Quantum chemistry\cite{bib:4831} 
is believed to be one of the most suitable research fields for quantum computation since its problem setting is quantum mechanical by definition.
Indeed,
a quantum computer can treat a many-electron state composed of lots of Slater determinants as it is in a sense that
the electronic state is encoded as a superposition of qubit states via an appropriately chosen map such as the Jordan--Wigner (JW)\cite{bib:4710} and Bravyi--Kitaev (BK)\cite{bib:4293} transformations.

The quantity which a quantum chemistry calculation is asked to first provide is the total energy of a target system.\cite{bib:4751}
One of the most widespread methods for obtaining the total energy is the variational quantum eigensolver (VQE),
in which a trial many-electron state is prepared via a parametrized quantum circuit.
The parameters are optimized iteratively with the aid of a classical computer aiming at the ground state.
This approach was first realized\cite{bib:4470} by using a quantum photonic device,
after which the realizations by superconducting\cite{bib:4289, bib:4292} and ion trap\cite{bib:4518} quantum computers have been reported.
There exist algorithms for obtaining the energy spectra of excited states.\cite{bib:4815, bib:4936, bib:4739, bib:4816, bib:4817, bib:4797}

Not only academic interest
but also industrial demands for accurate explanations and predictions of material properties make it an urgent task to develop methodologies on quantum computers for various electronic properties other than energy levels.
The one-particle Green's functions (GFs) are important in correlated electronic systems\cite{bib:4070, bib:4575}
since they are often used in explanation of spectra measured in photoemission and its inverse experiments.
Recently we proposed\cite{2019arXiv190803902K} a method for the GFs\cite{bib:4611, bib:4617} via statistical samplingby employing quantum phase estimation (QPE).\cite{Nielsen_and_Chuang}
Endo et al.,\cite{bib:4916} on the other hand,
proposed a method for the GFs by focusing on noisy intermediate-scale quantum (NISQ) devices.

% importance of response functions
The charge and spin response functions,
formulated in the linear-response theory,\cite{stefanucci2013nonequilibrium, bib:4276}
describe the leading contributions to the electric and magnetic excitations when perturbation fields are applied to a target system.
Since the response functions are the fundamental building blocks in constructing the elaborated methods for correlated electrons such as $GW$ theory,\cite{bib:GW1, stefanucci2013nonequilibrium}
the accurate calculation of them is needed.
In addition,
there exist response functions directly related to measurable quantities such as dielectric constants,
electric conductivities,
and magnetic susceptibilities.

% what we do in this study
Given the recent rapid development of fabrication techniques for quantum hardware and
the growing demands for quantum computation in material science,
it is worth making tools for analyses on correlation effects.
In this study,
we propose a scheme for the construction of the response functions of an interacting electronic system via statistical sampling on a quantum computer.
For examining the validity of our scheme,
we perform simulations of such construction of the response functions of diatomic molecules by referring to the full configuration-interaction (FCI) results.

% organization of this paper
This paper is organized as follows.
In Section \ref{sec:methods},
we explain the basic ideas and our scheme in detail by providing the quantum circuits for obtaining the response functions via statistical sampling.
In Section \ref{sec:computational_details},
we describe the computational details for our simulations on a classical computer.
In Section \ref{sec:results_and_discussion},
we show the simulation results for C$_2$ and N$_2$ molecules.
In Section \ref{sec:conclusions},
we provide the conclusions.

\section{Methods}
\label{sec:methods}

\subsection{Circuit for a linear combination of unitaries}

The calculation of a physical quantities of interest quite often involves the evaluation of matrix elements of various electronic operators.
Such an operator is, however,
not necessarily unitary and it prohibits one from implementing it straightforwardly as logic gates for qubits.
We describe a workaround for this difficulty by providing a circuit for probabilistic state preparation.

For $2^n$ unitaries $U_k$,
where $k$ is a bit string of length $n$,
we want to apply an arbitrary linear combination of them
\begin{gather}
    \mathcal{O}
    =
        c_{0 \cdots 00}
        U_{0 \cdots 00}
        +
        c_{0 \cdots 01}
        U_{0 \cdots 01}
        +
        \cdots
        +
        c_{1 \cdots 11}
        U_{1 \cdots 11}
    \label{def_lin_combo_unitaries}
\end{gather}
to an input register $| \psi \rangle$.
We can assume that all the coefficients $c$ are real and
$c_{0 \cdots 00} = 1$ without loss of generality.
We construct a circuit $\mathcal{C}_{\mathcal{O}}$ equipped with $n$ ancillary qubits,
as depicted in Fig. \ref{circuit_for_lin_combo_unitaries},
containing a partial circuit $\mathcal{C}^{(n)}$ defined recursively in Fig. \ref{part_circ_for_lin_combo_unitaries}. 
If the measurement outcome for the ancillae is $| 0 \rangle^{\otimes n}$,
the state of whole system collapses to $| 0 \rangle^{\otimes n} \otimes \mathcal{O} | \psi \rangle$ up to a normalization constant.
The proof of this fact is provided in Appendix \ref{appendix:proof_prob_stat_prep}.

The state preparation techniques for the response functions described below are spacial cases of the one introduced above.

\begin{figure}
\centering
\mbox{ 
\Qcircuit @C=1em @R=1em { 
    \lstick{| q^{\mathrm{A}}_{n-1} = 0 \rangle} & \gate{H} & \multigate{4}{\mathcal{C}^{(n)}} & \meter & \cw \\
                                                & \vdots   & \pureghost{\mathcal{C}^{(n)}}       & \vdots &  \\
    \lstick{| q^{\mathrm{A}}_1     = 0 \rangle} & \gate{H} &  \ghost{\mathcal{C}^{(n)}}       & \meter & \cw \\
    \lstick{| q^{\mathrm{A}}_0 = 0 \rangle}     & \gate{H} &  \ghost{\mathcal{C}^{(n)}}       & \meter & \cw \\
    \lstick{| \psi    \rangle}                  & {/} \qw  & \ghost{\mathcal{C}^{(n)}}        & \qw    & \qw  
} 
} 
\caption{
Circuit $\mathcal{C}_{\mathcal{O}}$ for the probabilistic preparation of a state on which a linear combination $\mathcal{O}$ of $2^n$ unitaries have acted.
$H$ in the circuit represents the Hadamard gate.
The desired state is prepared according to a measurement outcome of $n$ ancillary qubits.
This circuit contains $\mathcal{C}^{(n)}$ as a partial circuit defined in Fig. \ref{part_circ_for_lin_combo_unitaries}.
}
\label{circuit_for_lin_combo_unitaries} 
\end{figure}
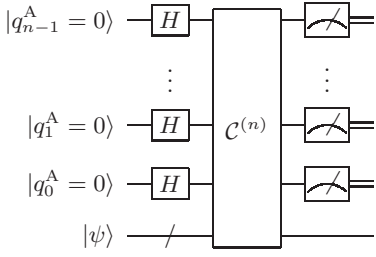

\begin{figure*}
\centering

\subfloat[For $j = 0$]{
\Qcircuit @C=1em @R=1em {
    {} & \qw & {/} \qw & \gate{\mathcal{C}^{(j=0)}_{\lambda}}  & \qw & \qw & \push{\equiv\rule{1em}{0em}} & {/} \qw & \gate{U_{\lambda}} & \qw & \qw
}
}

\subfloat[For $j > 0$]{
\Qcircuit @C=1em @R=1em {
   {} & \qw &     \qw & \multigate{2}{\mathcal{C}^{(j>0)}_{\lambda}} & \qw & \qw & \push{\rule{2em}{0em}}       &     \qw & \ctrlo{1}                                      & \ctrl{1}                                       & \gate{R_y \left(-2 \theta^{(j)}_{\lambda} \right)} & \qw \\
    {} & \ustick{\otimes (j-1)} \qw & {/} \qw & \ghost{\mathcal{C}^{(j>0)}_{\lambda}}        & \qw & \qw & \push{\equiv\rule{1em}{0em}} & {/} \qw & \multigate{1}{\mathcal{C}^{(j-1)}_{\lambda 0}} & \multigate{1}{\mathcal{C}^{(j-1)}_{\lambda 1}} & \qw & \qw \\
    {} & \qw                        & {/} \qw & \ghost{\mathcal{C}^{(j>0)}_{\lambda}}        & \qw & \qw & \push{\rule{2em}{0em}}       & {/} \qw & \ghost{\mathcal{C}^{(j-1)}_{\lambda 0}}        & \ghost{\mathcal{C}^{(j-1)}_{\lambda 1}}        & \qw & \qw  
}
}
\\
\caption{
Recursive definitions of circuits $\mathcal{C}_{\lambda}^{(j)}$ for
(a) $j = 0$ and (b) $j > 0$.
$\lambda$ is empty or a bit string.
$R_y (\theta) = e^{-i \theta \sigma_y/2}$ is a rotation for an angle $\theta$.
These circuits are used in Fig. \ref{circuit_for_lin_combo_unitaries}.
} 
\label{part_circ_for_lin_combo_unitaries} 
\end{figure*}
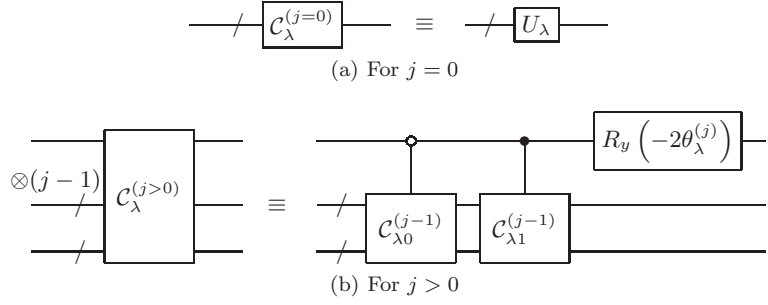

\subsection{Definitions for linear responses}

\subsubsection{Linear-response functions}

We work with the $n_{\mathrm{orbs}}$ orthonormalized spatial one-electron orbitals for each spin direction in a target $N$-electron system,
which we assume is in the many-electron ground state $| \Psi_{\mathrm{gs}} \rangle$ at zero temperature.
The formalism described below can be easily extended to a system with multiple ground states and/or a nonzero temperature.

The response functions in terms of Hermitian operators $\mathcal{O}$ and $\mathcal{O}'$ in time domain is defined as\cite{stefanucci2013nonequilibrium}
\begin{gather}
    \chi_{\mathcal{O O'}} (t, t')
    \equiv
        -i
        \theta (t -t')
        \langle
        [
            \mathcal{O} (t),
            \mathcal{O}' (t')
        ]
        \rangle
    ,
    \label{def_spin_resp_func}
\end{gather}
where
$\mathcal{O} (t)$ is the operator in the Heisenberg picture
and the expectation value is for the ground state.
We assume that the system has been in the equilibrium when the perturbation is turned on.
Since the response function defined in eq. (\ref{def_spin_resp_func})
depends on time only via $t - t'$ in this case,
its expression in frequency domain is written as
\begin{gather}
    \chi_{\mathcal{O O'}} (\omega)
    =
        R_{\mathcal{O O'}} (\omega + i \delta)
        +
        R_{\mathcal{O' O}} (-\omega - i \delta)
    \label{resp_func_using_R}
\end{gather}
for a real frequency $\omega$.
\begin{gather}
    R_{\mathcal{O O'}} (z)
    \equiv
        \sum_{\lambda}
            \frac{L_{\lambda \mathcal{O O'}}}{z - (E_{\lambda} - E^N_{\mathrm{gs}}) }
    \label{def_R_for_resp}
\end{gather}
is the Lehmann summation over the energy eigenstates
for a complex frequency $z$.
$E^N_{\mathrm{gs}}$ is the ground-state energy and
$E_{\lambda}$ is the energy eigenvalue of the $\lambda$th excited state $| \Psi_{\lambda} \rangle$.
\begin{gather}
    L_{\lambda \mathcal{O O'}}
    \equiv
        \langle \Psi_{\mathrm{gs}} |
        \mathcal{O} 
        | \Psi_{\lambda} \rangle
        \langle \Psi_{\lambda} |   
        \mathcal{O}' 
        | \Psi_{\mathrm{gs}} \rangle
    \label{def_generic_trans_mat}
\end{gather}
is the transition matrix element,
which satisfies clearly
$
L_{\lambda \mathcal{O O'}}
=
(L_{\lambda \mathcal{O' O}})^*
.
$
The positive infinitesimal constant $\delta$ appears in eq. (\ref{resp_func_using_R}) due to the retarded nature of the response function,
rendering all the poles immediately below the real axis.
It is clear that the real part of $\chi_{\mathcal{O O'}} (\omega)$ is even with respect to $\omega$,
while the imaginary part is odd.

\subsubsection{Charge and spin responses}

By using the creation $a_{p \sigma}^\dagger$ and
annihilation $a_{p \sigma}$ operators
for the $p$th spatial orbital of $\sigma$ spin
$| \phi_{p \sigma} \rangle$
$(\sigma = \alpha, \beta)$,
the electron number operator is given by
$n_{p \sigma} = a_{p \sigma}^\dagger a_{p \sigma}$.
The spin operator is given by
$
\boldsymbol{s}_p
=
\sum_{\sigma, \sigma'}
a_{p \sigma}^\dagger
(\boldsymbol{\sigma}^{\mathrm{el}}_{\sigma \sigma'} /2)
a_{p \sigma'}
,
$
where $\boldsymbol{\sigma}^{\mathrm{el}}$ is the Pauli matrix for the electronic state.

The orbital-wise response functions involving the charge and spin of the individual spin orbitals are obtained by putting
$\mathcal{O}_{p n} \equiv \sum_{\sigma} n_{p \sigma}$
and
$\mathcal{O}_{p j} \equiv s_{p j} \, (j = x, y, z)$,
respectively,
into eq. (\ref{def_R_for_resp}).
To rewrite the expression for transition matrix elements
in eq. (\ref{def_generic_trans_mat}) into a more tractable form,
we define the charge-charge transition matrix element
\begin{gather}
    N_{\lambda p \sigma, p' \sigma'}
    \equiv
        \langle \Psi_{\mathrm{gs}} |
        n_{p \sigma} 
        | \Psi_{\lambda} \rangle
        \langle \Psi_{\lambda} |   
        n_{p' \sigma'} 
        | \Psi_{\mathrm{gs}} \rangle
    \label{def_transition_mat_for_occ}
\end{gather}
for the $\lambda$th energy eigenstate $| \Psi_{\lambda} \rangle$ of the $N$-electron states.
We define similarly the spin-spin one
\begin{gather}
    S_{\lambda p j, p' j'}
    \equiv
        \langle \Psi_{\mathrm{gs}} |
        s_{p  j} 
        | \Psi_{\lambda} \rangle
        \langle \Psi_{\lambda} |   
        s_{p'  j'} 
        | \Psi_{\mathrm{gs}} \rangle
    \label{def_transition_mat_for_spin}
\end{gather}
for $j, j' = x, y, z$ and the spin-charge one
\begin{gather}
    M_{\lambda p j, p' \sigma'}
    \equiv
        \langle \Psi_{\mathrm{gs}} |
        s_{p j} 
        | \Psi_{\lambda} \rangle
        \langle \Psi_{\lambda} |   
        n_{p' \sigma'} 
        | \Psi_{\mathrm{gs}} \rangle
    \equiv
        M_{\lambda p' \sigma', p j}^*
    .
    \label{def_transition_mat_for_spin_occ}
\end{gather}
From these matrix elements,
we define the following Hermitian matrices $L_{\lambda}$:
\begin{gather}
    L_{\lambda p n, p' n}
    \equiv
        \sum_{\sigma, \sigma'}
            N_{\lambda p \sigma, p' \sigma'}
    \\
    L_{\lambda p j, p' j'}
    \equiv
        S_{\lambda p j, p' j'}
    \\
    L_{\lambda p j, p' n}
    \equiv
        \sum_{\sigma}
            M_{\lambda p j, p' \sigma}
    ,
\end{gather}
which are used for eq. (\ref{def_generic_trans_mat}).

\subsubsection{Generic one-body operators}

A generic one-body operator is given in the form
\begin{gather}
    \mathcal{O}
    =
        \sum_{m, m'}
            \mathcal{O}_{m m'}
            a_{m}^\dagger
            a_{m'}
    ,
    \label{resp_one_body_def_one_body_opr}
\end{gather}
where $m$ and $m'$ are the composite indices of spatial orbitals and spins.
$\mathcal{O}_{m m'}$ is the matrix element between the one-electron orbitals.
Our scheme is actually applicable to such a generic case for $\mathcal{O}$ and $\mathcal{O}'$,
for which we have to obtain the transition matrix elements
\begin{gather}
    B_{\lambda m_1 m_2, m_3 m_4}
    \equiv
	    \langle \Psi_{\mathrm{gs}} |
	    a_{m_1}^\dagger
	    a_{m_2}
	    | \Psi_\lambda \rangle
	    \langle \Psi_\lambda |
	    a_{m_3}^\dagger
	    a_{m_4}
	    | \Psi_{\mathrm{gs}} \rangle
    ,
    \label{resp_one_body_def_transition_mat}
\end{gather}
which satisfy clearly
$
B_{\lambda m_1 m_2, m_3 m_4}
=
(B_{\lambda m_4 m_3, m_2 m_1})^*
.
$
From them,
we sum up the contributions to eq. (\ref{def_generic_trans_mat}) as 
\begin{gather}
    L_{\lambda \mathcal{O O'}}
    =
        \sum_{m_1, m_2, m_3, m_4}
            \mathcal{O}_{m_1 m_2}
            \mathcal{O}_{m_3 m_4}'
            B_{\lambda m_1 m_2, m_3 m_4}
        .
    \label{resp_one_body_sum_of_transition_mats}
\end{gather}

We calculate the electric polarizabilities of molecules in the present study as examples for the generic scheme.
The contribution from electrons to the electric dipole of a molecule is
$
    \boldsymbol{d}
    =
        -
        \sum_{m, m'}
            \boldsymbol{d}_{m m'}
            a_{m}^\dagger
            a_{m'}
        ,
$
where the negative sign on the right-hand side comes from the electron charge.
$
\boldsymbol{d}_{m m'}
\equiv
\langle \phi_m | \boldsymbol{r} | \phi_{m'} \rangle
$
is the matrix element of position operator.
The linear electric polarizability tensor $\alpha_{j j'} \, (j, j' = x, y, z)$\cite{bib:4988, bib:4989} is defined as the first derivative of the expected electric dipole with respect to an external electric field,
obtained via the response function as
$
    \alpha_{j j'} (\omega)
    =
        -
        \chi_{d_j d_{j'}} (\omega)
        ,
$
from which the photoabsorption cross section\cite{bib:4616} is given by
\begin{gather}
    \sigma (\omega)
    =
        \frac{4 \pi}{c}
        \omega
        \mathrm{Im Tr} \,
        \alpha (\omega)
        .
    \label{def_photoabs_cross_section}
\end{gather}
This cross section is a measurable quantity in optical absorption experiments.

\subsubsection{Unitary decomposition of electronic operators}

Although there are alternatives for mapping the electronic operators of a target system to the qubit ones such as JW\cite{bib:4710} and BK\cite{bib:4293} transformations,
we do not distinguish between an electronic operator and
its corresponding qubit representation in what follows
since no confusion will occur for the readers.

For each combination of a spatial orbital $p$ and a spin $\sigma$,
we perform the Majorana fermion-like\cite{bib:4979} transformation for the qubits:\cite{2019arXiv190803902K}
\begin{gather}
    U_{0 p \sigma}
    =
        a_{p \sigma}
        +
        a_{p \sigma}^\dagger
    \label{unitary_oprs_using_el_0}
\end{gather}
and
\begin{gather}
    U_{1 p \sigma}
    =
        a_{p \sigma}
        -
        a_{p \sigma}^\dagger
    ,
    \label{unitary_oprs_using_el_1}
\end{gather}
which are unitary regardless of the adopted qubit representation
thanks to the anti-commutation relation between the electronic operators and
can thus be implemented as logic gates in the quantum computer.
This means that 
we can prepare at least probabilistically an electronic state on which an arbitrary product of the creation and annihilation operators has acted,
similarly to the case for GFs.\cite{2019arXiv190803902K}

In what follows,
we assume that the many-electron ground state $| \Psi_{\mathrm{gs}} \rangle$ is already known and can be prepared on a quantum computer.

\subsection{Charge-charge responses}
\label{subsec:chrage_charge_contr}

Let us first consider the determination of the charge-charge transition matrices $N_\lambda$.

\subsubsection{Circuits for diagonal components}

From the unitary operators in eqs. (\ref{unitary_oprs_using_el_0}) and (\ref{unitary_oprs_using_el_1}) for a combination of
a spatial orbital $p$ and a spin $\sigma$,
we define
$
    U^{(p)}_{\kappa \kappa' \sigma}
    \equiv
        U_{\kappa p \sigma}
        U_{\kappa' p \sigma}
$
for $\kappa, \kappa' = 0, 1$,
which are also unitary.
With them, the electron number operator is written as
\begin{gather}
    n_{p \sigma}
    =
        \frac{U^{(p)}_{0 0 \sigma} + U^{(p)}_{0 1 \sigma} - U^{(p)}_{1 0 \sigma} - U^{(p)}_{1 1 \sigma}}{4}
    ,
    \label{el_occ_opr_using_unitary}
\end{gather}
while the hole number operator is written as
\begin{gather}
    \widetilde{n}_{p \sigma}
    \equiv
        1 - n_{p \sigma}
    =
        \frac{U^{(p)}_{0 0 \sigma} - U^{(p)}_{0 1 \sigma} + U^{(p)}_{1 0 \sigma} - U^{(p)}_{1 1 \sigma}}{4}
    .
    \label{el_occ_opr_using_unitary_tilde}
\end{gather}

\begin{figure}
\centering
\mbox{ 
\Qcircuit @C=0.8em @R=1em { 
                                                                &          & \mathcal{U}^{(p)}_{\sigma}  &            &            &            &          &        & \\        
    \lstick{| q^{\mathrm{A}}_1 = 0 \rangle} & \gate{H} & \ctrlo{1}  & \ctrlo{1}  & \ctrl{1}   & \ctrl{1}   & \gate{H} & \qw & \qw \\
    \lstick{| q^{\mathrm{A}}_0 = 0 \rangle} & \gate{H} & \ctrlo{1}  & \ctrl{1}   & \ctrlo{1}  & \ctrl{1}   & \gate{H} & \meter & \cw \\
    \lstick{| \psi \rangle}       & {/} \qw & \gate{U^{(p)}_{0 0 \sigma}} & \gate{U^{(p)}_{0 1 \sigma}}    & \gate{U^{(p)}_{1 0 \sigma}} & \gate{U^{(p)}_{1 1 \sigma}} & \qw      & \qw    & \rstick{| \widetilde{\psi} \rangle} \qw \gategroup{2}{3}{4}{6}{1em}{--}
} 
} 
\caption{
Charge-charge diagonal circuit
$\mathcal{C}_{p \sigma} \, (\sigma = \alpha, \beta)$
for probabilistic preparation of
$n_{p \sigma} | \psi \rangle$ and
$\widetilde{n}_{p \sigma} | \psi \rangle$
from an arbitrary input state $| \psi \rangle$ and
two ancillary qubits.
We define the partial circuit $\mathcal{U}^{(p)}_{\sigma}$ by enclosing it with dashed lines.
} 
\label{circuit_prep_el_occ_alpha_beta} 
\end{figure}
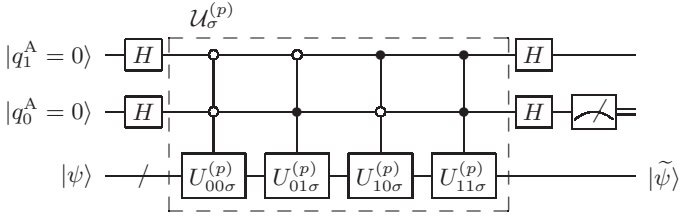

We construct a circuit $\mathcal{C}_{p \sigma}$ equipped with two ancillary qubits $| q_0^{\mathrm{A}} \rangle$ and $| q_1^{\mathrm{A}} \rangle$ by implementing the controlled operations of $U^{(p)}_{\kappa \kappa' \sigma}$,
as depicted in Fig. \ref{circuit_prep_el_occ_alpha_beta}. 
The whole system consists of the ancillae and an arbitrary input register $| \psi \rangle$.
Its state changes by undergoing the circuit as
\begin{gather}
    | q^{\mathrm{A}}_1 = 0 \rangle
    \otimes
    | q^{\mathrm{A}}_0 = 0 \rangle
    \otimes
    | \psi \rangle
    \nonumber \\
    \longmapsto
        | 0 \rangle \otimes
        | 1 \rangle \otimes
        \widetilde{n}_{p \sigma} | \psi \rangle
        +
        | 1 \rangle \otimes
        | 0 \rangle \otimes
        n_{p \sigma} | \psi \rangle
    \equiv
        | \Phi_{p \sigma} \rangle
    ,
    \label{state_trans_occ_diag}
\end{gather}
since $(a_{p \sigma}^{\dagger})^2$ and $a_{p \sigma}^2$ vanish due to the Fermi statistics.
The projective measurement\cite{Nielsen_and_Chuang} on $| q_0^{\mathrm{A}} \rangle$ is represented by the two operators
$\mathcal{P}_q = I \otimes | q \rangle \langle q | \otimes I \, (q = 0, 1)$.
The state of the whole system collapses immediately after the measurement as follows:
\begin{gather}
    | \Phi_{p \sigma} \rangle
    \overset{| 0 \rangle \, \mathrm{observed}}{\longmapsto}
        | 1 \rangle
        \otimes
        | 0 \rangle
        \otimes
        \frac{n_{p \sigma}}{\sqrt{\mathbb{P}_{p \sigma}}}
        | \psi \rangle
    \nonumber \\
    \mathrm{prob.}
    \,
        \left\| n_{p \sigma} | \psi \rangle \right\|^2
    \equiv
        \mathbb{P}_{p \sigma}
    \label{state_trans_occ_occ}
    ,
    \\
    | \Phi_{p \sigma} \rangle
    \overset{| 1 \rangle \, \mathrm{observed}}{\longmapsto}
        | 0 \rangle
        \otimes
        | 1 \rangle
        \otimes
        \frac{\widetilde{n}_{p \sigma}}{\sqrt{\widetilde{\mathbb{P}}_{p \sigma}}}
        | \psi \rangle
    \nonumber \\
    \mathrm{prob.}
    \,
        \left\| \widetilde{n}_{p \sigma} | \psi \rangle \right\|^2
    \equiv
        \widetilde{\mathbb{P}}_{p \sigma}
    .
\end{gather}

\subsubsection{Circuits for off-diagonal components}

For mutually different combinations $(p, \sigma)$ and $(p', \sigma')$ of spatial orbitals and spins,
we define the four non-Hermitian auxiliary operators
\begin{gather}
    n_{p \sigma, p' \sigma'}^\pm
    \equiv
        \frac{
            n_{p \sigma}
            \pm
            e^{i \pi/4}
            n_{p' \sigma'}
        }{2}
    \label{def_opr_occ_aux_el}
\end{gather}
and
\begin{gather}
    \widetilde{n}_{p \sigma, p' \sigma'}^\pm
    \equiv
        \frac{
            \widetilde{n}_{p \sigma}
            \pm
            e^{i \pi/4}
            \widetilde{n}_{p' \sigma'}
        }{2}
    .
    \label{def_opr_occ_aux_hole}
\end{gather}
Unnormalized auxiliary states
$
| \Psi_{p \sigma, p' \sigma'}^\pm \rangle
\equiv
n_{p \sigma, p' \sigma'}^\pm
| \Psi_{\mathrm{gs}} \rangle
$
can have overlaps
$
T_{\lambda p \sigma, p' \sigma'}^\pm
\equiv
| \langle \Psi_\lambda | \Psi_{p \sigma, p' \sigma'}^\pm \rangle |^2
$
with the energy eigenstates,
from which the charge-charge transition matrix elements in eq. (\ref{def_transition_mat_for_occ}) can be calculated as
\begin{gather}
    N_{\lambda p \sigma, p' \sigma'}
    =
        e^{-i \pi/4}
        (
            T_{\lambda p \sigma, p' \sigma'}^+
            -
            T_{\lambda p \sigma, p' \sigma'}^-
        )
    \nonumber \\
        +
        e^{i \pi/4}
        (
            T_{\lambda p' \sigma', p \sigma}^+
            -
            T_{\lambda p' \sigma', p \sigma}^-
        )
    .
    \label{transition_occ_off_diag_alpha_half}
\end{gather}

\begin{figure}
\centering
\mbox{ 
\Qcircuit @C=0.8em @R=1em { 
    \lstick{| q^{\mathrm{A}}_2 = 0 \rangle} & \gate{H} & \ctrlo{1}                                 & \ctrl{1}                                    & \gate{Z \left( \frac{\pi}{4} \right)} & \gate{H} & \meter & \cw \\
    \lstick{| q^{\mathrm{A}}_1 = 0 \rangle} & \gate{H} & \multigate{2}{\mathcal{U}^{(p)}_{\sigma}} & \multigate{2}{\mathcal{U}^{(p')}_{\sigma'}} & \qw                                   & \gate{H} & \meter & \cw \\
    \lstick{| q^{\mathrm{A}}_0 = 0 \rangle} & \gate{H} & \ghost{\mathcal{U}^{(p)}_{\sigma}}        & \ghost{\mathcal{U}^{(p')}_{\sigma'}}        & \qw                                   & \gate{H} & \qw & \qw \\
    \lstick{| \psi \rangle}                 & {/} \qw  & \ghost{\mathcal{U}^{(p)}_{\sigma}}        & \ghost{\mathcal{U}^{(p')}_{\sigma'}}        & \qw                                   & \qw & \qw & \rstick{| \widetilde{\psi} \rangle} \qw
} 
} 
\caption{
Charge-charge off-diagonal circuit
$\mathcal{C}_{p \sigma, p' \sigma'} \, (\sigma, \sigma' = \alpha, \beta)$
for probabilistic preparation of
$n_{p \sigma, p' \sigma'}^\pm | \psi \rangle$
and
$\widetilde{n}_{p \sigma, p' \sigma'}^\pm | \psi \rangle$
from an arbitrary input state $| \psi \rangle$ and three ancillary qubits.
$Z (\pm \pi/4) = \mathrm{diag}(1, e^{\pm i \pi/4})$ is a phase gate.
The partial circuits defined in Fig. \ref{circuit_prep_el_occ_alpha_beta} are contained as the controlled subroutines.
} 
\label{circuit_prep_el_occ_alpha_beta_aux_using_U} 
\end{figure}
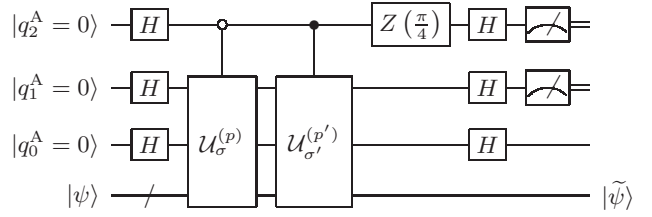

We construct a circuit $\mathcal{C}_{p \sigma, p' \sigma'}$ equipped with three ancillary qubits by using the controlled operations of the partial circuits $\mathcal{U}^{(p)}_{\sigma}$ and $\mathcal{U}^{(p')}_{\sigma'}$,
defined in Fig. \ref{circuit_prep_el_occ_alpha_beta},
as depicted in Fig. \ref{circuit_prep_el_occ_alpha_beta_aux_using_U}. 
The whole system consists of the ancillae and an arbitrary input register $| \psi \rangle$.
Its state changes by undergoing the circuit as
\begin{gather}
    | q^{\mathrm{A}}_2 = 0 \rangle
    \otimes
    | q^{\mathrm{A}}_1 = 0 \rangle
    \otimes
    | q^{\mathrm{A}}_0 = 0 \rangle
    \otimes
    | \psi \rangle
    \nonumber \\
    \longmapsto
        | 0 \rangle \otimes
        | 0 \rangle \otimes
        | 1 \rangle \otimes
        \widetilde{n}_{p \sigma, p' \sigma'}^+
        | \psi \rangle
    \nonumber \\
        +
        | 0 \rangle \otimes
        | 1 \rangle \otimes
        | 0 \rangle \otimes
        n_{p \sigma, p' \sigma'}^+
        | \psi \rangle
    \nonumber \\
        +
        | 1 \rangle \otimes
        | 0 \rangle \otimes
        | 1 \rangle \otimes
        \widetilde{n}_{p \sigma, p' \sigma'}^-
        | \psi \rangle
    \nonumber \\
        +
        | 1 \rangle \otimes
        | 1 \rangle \otimes
        | 0 \rangle \otimes
        n_{p \sigma, p' \sigma'}^-
        | \psi \rangle
    \nonumber \\
    \equiv
        | \Phi_{p \sigma, p' \sigma'} \rangle
    .
    \label{state_trans_occ_off_diag}
\end{gather}
The projective measurement on $| q^{\mathrm{A}}_2 \rangle$ and $| q^{\mathrm{A}}_1 \rangle$ is represented by the four operators
$\mathcal{P}_{q q'} = | q \rangle \langle q | \otimes | q' \rangle \langle q' | \otimes I \otimes I \, (q, q' = 0, 1)$.
The two outcomes among the possible four are of our interest,
immediately after which the whole system collapses as follows:
\begin{gather}
    | \Phi_{p \sigma, p' \sigma'} \rangle
    \overset{| 0 \rangle \otimes | 1 \rangle \, \mathrm{observed}}{\longmapsto}
        | 0 \rangle
        \otimes
        | 1 \rangle
        \otimes
        | 0 \rangle
        \otimes
        \frac{n_{p \sigma, p' \sigma'}^+ }{\sqrt{\mathbb{P}_{p \sigma, p' \sigma'}^+}}
        | \psi \rangle
    \nonumber \\
    \mathrm{prob.}
    \,
        \left\| n_{p \sigma, p' \sigma'}^+ | \psi \rangle \right\|^2
    \equiv
        \mathbb{P}_{p \sigma, p' \sigma'}^+
    ,
    \label{state_trans_occ_occ_off_diag_0_1}
    \\
    | \Phi_{p \sigma, p' \sigma'} \rangle
    \overset{| 1 \rangle \otimes | 1 \rangle \, \mathrm{observed}}{\longmapsto}
        | 1 \rangle
        \otimes
        | 1 \rangle
        \otimes
        | 0 \rangle
        \otimes
        \frac{n_{p \sigma, p' \sigma'}^- }{\sqrt{\mathbb{P}_{p \sigma, p' \sigma'}^-}}
        | \psi \rangle
    \nonumber \\
    \mathrm{prob.}
    \,
        \left\| n_{p \sigma, p' \sigma'}^- | \psi \rangle \right\|^2
    \equiv
        \mathbb{P}_{p \sigma, p' \sigma'}^-
    .
    \label{state_trans_occ_occ_off_diag_1_1}
\end{gather}

\subsubsection{Transition matrices via statistical sampling}

Given the result of a measurement on the ancillary bits for a diagonal or an off-diagonal component,
we have the register $| \widetilde{\psi} \rangle$ different from the input $N$-electron state.
Then we perform QPE for the Hamiltonian $\mathcal{H}$ by inputting $| \widetilde{\psi} \rangle$ to obtain one of the energy eigenvalues in the Hilbert subspace for the $N$-electron states.
A QPE experiment inevitably suffers from probabilistic errors that depend on the number of qubits and 
the various parameters for the Suzuki--Trotter decomposition of $\mathcal{H}$.
We assume for simplicity, however, that the QPE procedure is realized on a quantum computer with ideal precision as well as in our previous study.\cite{2019arXiv190803902K}
We will thus find the estimated value to be $E^N_{\lambda}$ with a probability
$| \langle \Psi_\lambda | \widetilde{\psi} \rangle |^2$.
\cite{Nielsen_and_Chuang}

If we input $| \Psi_{\mathrm{gs}} \rangle$
to the diagonal circuit $\mathcal{C}_{p \sigma}$
in Fig. \ref{circuit_prep_el_occ_alpha_beta} and
observe the ancillary bit $| q^{\mathrm{A}}_0 = 0 \rangle$ for QPE,
the energy eigenvalue $E^N_{\lambda}$ will be obtained with a probability
[see eq. (\ref{state_trans_occ_occ})]
\begin{gather}
        \left|
        \langle \Psi_\lambda |
        \frac{n_{p \sigma}}{\sqrt{\mathbb{P}_{p \sigma}}}
        | \Psi_{\mathrm{gs}} \rangle
        \right|^2
        \mathbb{P}_{p \sigma}
    =
        N_{\lambda p \sigma, p \sigma}
    .
    \label{prob_energy_occ_occ_diag}
\end{gather}
This means that we can get the diagonal components of transition matrices $N_{\lambda}$ via statistical sampling for a fixed combination of $p$ and $\sigma$.

If we input $| \Psi_{\mathrm{gs}} \rangle$ to the off-diagonal circuit $\mathcal{C}_{p \sigma, p' m'}$ in Fig. \ref{circuit_prep_el_occ_alpha_beta_aux_using_U} and
observe the ancillary bits
$
| q^{\mathrm{A}}_2 = 0 \rangle
\otimes
| q^{\mathrm{A}}_1 = 1 \rangle
$
or
$
| q^{\mathrm{A}}_2 = 1 \rangle
\otimes
| q^{\mathrm{A}}_1 = 1 \rangle
$
for QPE,
the energy eigenvalue $E^N_{\lambda}$ will be obtained with probabilities
[see eqs. (\ref{state_trans_occ_occ_off_diag_0_1}) and (\ref{state_trans_occ_occ_off_diag_1_1})]
\begin{gather}
        \left|
        \langle \Psi_\lambda |
        \frac{n_{p \sigma, p' \sigma'}^{\pm} }{\sqrt{\mathbb{P}_{p \sigma, p' \sigma'}^{\pm}}}
        | \Psi_{\mathrm{gs}} \rangle
        \right|^2
        \mathbb{P}_{p \sigma, p' \sigma'}^{\pm}
    =
        T_{\lambda p \sigma, p' \sigma'}^{\pm}
    .
    \label{prob_energy_occ_occ_off_diag}
\end{gather}
This means that we can get the off-diagonal components of $N_{\lambda
}$
from eq. (\ref{transition_occ_off_diag_alpha_half})
via statistical sampling for a fixed combination of $p, p', \sigma$, and $\sigma'$.

\subsection{Spin-spin responses}
\label{subsec:spin_spin_contr}

Let us next consider the determination of the spin-spin transition matrix elements $S_{\lambda p j, p' j'}$ for $j, j' = x, y$.
Those involving the $z$ components of spins can be calculated
from $N_{\lambda}$ by using the relation
$s_{p z} = (n_{p \alpha} - n_{p \beta})/2$.

\subsubsection{Circuits for diagonal components}

From the unitary operators in eqs. (\ref{unitary_oprs_using_el_0}) 
and (\ref{unitary_oprs_using_el_1}) for a combination of
a spatial orbital $p$ and a spin $\sigma$,
we define the following four unitary operators:
$
    U^{(p)}_{0 x}
    \equiv
        U_{0 p \alpha} U_{1 p \beta}
    , \,
    U^{(p)}_{1 x}
    \equiv
        -U_{1 p \alpha} U_{0 p \beta}
    , \,
    U^{(p)}_{0 y}
    \equiv
        -i
        U_{0 p \alpha} U_{0 p \beta}
    ,
$        
and
$
    U^{(p)}_{1 y}
    \equiv
        i
        U_{1 p \alpha} U_{1 p \beta}
    .
$
With them,
the spin operators for the $x$ and $y$ directions are written as
\begin{gather}
    s_{p  j}
    =
        \frac{U^{(p)}_{0 j} + U^{(p)}_{1 j}}{4}
    \label{el_spin_opr_using_unitary_x_y}
\end{gather}
for $j = x, y$.
We define 
\begin{gather}
    \widetilde{s}_{p j}
    \equiv
        \frac{U^{(p)}_{0 j} - U^{(p)}_{1 j}}{4}
    \label{el_spin_opr_using_unitary_tilde}
\end{gather}
for later convenience.

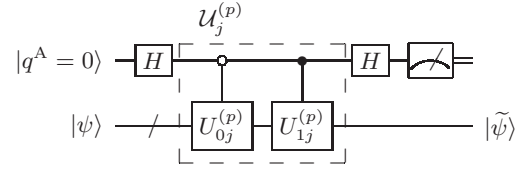
\begin{figure} 
\centering
\mbox{ 
\Qcircuit @C=0.8em @R=1em { 
                                                              &          & \mbox{$\mathcal{U}^{(p)}_j$}   &                      &          &           &   &     & \\ 
    \lstick{| q^{\mathrm{A}} = 0 \rangle} & \gate{H} & \ctrlo{1}            & \ctrl{1}             & \gate{H} & \meter    & \cw \\
    \lstick{| \psi \rangle}               & {/} \qw & \gate{U^{(p)}_{0 j}} & \gate{U^{(p)}_{1 j}} & \qw      & \qw       & \rstick{| \widetilde{\psi} \rangle} \qw \gategroup{2}{3}{3}{4}{1em}{--}
} 
} 
\caption{
Spin-spin diagonal circuit
$\mathcal{C}_{p j} \, (j = x, y)$
for probabilistic preparation of
$s_{p j} | \psi \rangle$ and
$\widetilde{s}_{p j} | \psi \rangle$
from an arbitrary input state $| \psi \rangle$ and
an ancillary qubit.
We define the partial circuit $\mathcal{U}^{(p)}_j$ by enclosing it with dashed lines.
} 
\label{circuit_prep_el_spin_x_y} 
\end{figure}

We construct a circuit $\mathcal{C}_{p j}$ equipped with an ancillary qubit by implementing the controlled operations of $U^{(p)}_{0 j}$ and $U^{(p)}_{1 j}$,
as depicted in Fig. \ref{circuit_prep_el_spin_x_y}. 
The whole system consists of the ancilla and an arbitrary input register $| \psi \rangle$.
Its state changes by undergoing the circuit as
\begin{gather}
    | q^{\mathrm{A}} = 0 \rangle 
    \otimes
    | \psi \rangle
    \longmapsto
        | 0 \rangle
        \otimes
        2 s_{p j}
        | \psi \rangle
        +
        | 1 \rangle
        \otimes
        2 \widetilde{s}_{p j}
        | \psi \rangle
    \nonumber \\
    \equiv
        | \Phi_{p j} \rangle
    .
\end{gather}
The projective measurement on $| q^{\mathrm{A}} \rangle$ is represented by the two operators
$\mathcal{P}_q = | q \rangle \langle q | \otimes I \, (q = 0, 1)$.
The state of the whole system collapses immediately after the measurement as follows:
\begin{gather}
    | \Phi_{p j} \rangle
    \overset{| 0 \rangle \, \mathrm{observed}}{\longmapsto}
        | 0 \rangle
        \otimes
        \frac{2 s_{p j}}{ \sqrt{\mathbb{P}_{p j} } }
        | \psi \rangle
    \nonumber \\
    \mathrm{prob.}
    \,
        \left\| 2 s_{p j} | \psi \rangle \right\|^2
    \equiv
        \mathbb{P}_{p j}
    ,
    \label{state_trans_spin_spin_diag}
    \\
    | \Phi_{p j} \rangle
    \overset{| 1 \rangle \, \mathrm{observed}}{\longmapsto}
        | 1 \rangle
        \otimes
        \frac{2 \widetilde{s}_{p j}}{ \sqrt{\widetilde{\mathbb{P}}_{p j} } }
        | \psi \rangle
    \nonumber \\
    \mathrm{prob.}
    \,
        \left\| 2 \widetilde{s}_{p j} | \psi \rangle \right\|^2
    \equiv
        \widetilde{\mathbb{P}}_{p j}
    .
\end{gather}

\subsubsection{Circuits for off-diagonal components}

For mutually different combinations $(p, j)$ and $(p', j')$ of spatial orbitals and spin components $j, j' = x, y$,
we define the four non-Hermitian auxiliary operators
\begin{gather}
    s_{p  j, p' j'}^\pm
    \equiv
        \frac{
            s_{p  j}
            \pm
            e^{i \pi/4}
            s_{p'  j'}
        }{2}
    \label{def_opr_spin_aux_1}
\end{gather}
and
\begin{gather}
    \widetilde{s}_{p  j, p' j'}^\pm
    \equiv
        \frac{
            \widetilde{s}_{p  j}
            \pm
            e^{i \pi/4}
            \widetilde{s}_{p'  j'}
        }{2}
    .
    \label{def_opr_spin_aux_2}
\end{gather}
Unnormalized auxiliary states
$
| \Psi_{p j, p' j'}^\pm \rangle
\equiv
s_{p  j, p' j'}^\pm
| \Psi_{\mathrm{gs}} \rangle
$
can have overlaps
$
T_{\lambda p j, p' j'}^\pm
\equiv
| \langle \Psi_\lambda | \Psi_{p j, p' j'}^\pm \rangle |^2
$
with the energy eigenstates,
from which the spin-spin transition matrix elements in eq. (\ref{def_transition_mat_for_spin}) can be calculated as
\begin{gather}
    S_{\lambda p j, p' j'}
    =
        e^{-i \pi/4}
        (
            T_{\lambda p j, p' j'}^+
            -
            T_{\lambda p j, p' j'}^-
        )
    \nonumber \\
        +
        e^{i \pi/4}
        (
            T_{\lambda p' j', p j}^+
            -
            T_{\lambda p' j', p j}^-
        )
    .
    \label{transition_spin_resp_off_diag_alpha_half}
\end{gather}

\begin{figure}
\centering
\mbox{ 
\Qcircuit @C=0.8em @R=1em { 
    \lstick{| q^{\mathrm{A}}_1 = 0 \rangle} & \gate{H} & \ctrlo{1}                          & \ctrl{1}                               & \gate{Z \left( \frac{\pi}{4} \right)} & \gate{H} & \meter & \cw \\
    \lstick{| q^{\mathrm{A}}_0 = 0 \rangle} & \gate{H} & \multigate{1}{\mathcal{U}^{(p)}_j} & \multigate{1}{\mathcal{U}^{(p')}_{j'}} & \qw                                   & \gate{H} & \meter & \cw \\
    \lstick{| \psi \rangle}                 & {/} \qw & \ghost{\mathcal{U}^{(p)}_j}         & \ghost{\mathcal{U}^{(p')}_{j'}}        & \qw                                   & \qw      & \qw    & \rstick{| \widetilde{\psi} \rangle} \qw
}  
} 
\caption{
Spin-spin off-diagonal circuit
$\mathcal{C}_{p j, p' j'} \, (j, j' = x, y)$
for probabilistic preparation of
$s_{p j, p' j'}^{\pm} | \psi \rangle$
and
$\widetilde{s}_{p j,p'  j'}^{\pm} | \psi \rangle$
from an arbitrary input state $| \psi \rangle$ and two ancillary qubits.
The partial circuits defined in Fig. \ref{circuit_prep_el_spin_x_y} are contained as the controlled subroutines.
} 
\label{circuit_prep_el_spin_x_y_aux_using_U} 
\end{figure}
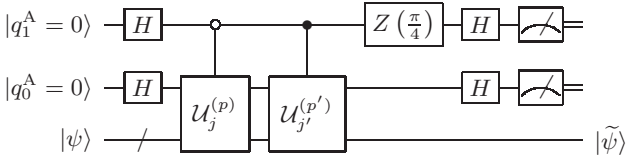

We construct a circuit $\mathcal{C}_{p j, p' j'}$ equipped with two ancillary qubits by using the controlled operations of the partial circuits $\mathcal{U}^{(p)}_{j}$ and $\mathcal{U}^{(p')}_{j'}$,
defined in Fig. \ref{circuit_prep_el_spin_x_y},
as depicted in Fig. \ref{circuit_prep_el_spin_x_y_aux_using_U}. 
The whole system consists of the ancillae and an arbitrary input register $| \psi \rangle$.
Its state changes by undergoing the circuit as
\begin{gather}
    | q^{\mathrm{A}}_1 = 0 \rangle
    \otimes
    | q^{\mathrm{A}}_0 = 0 \rangle
    \otimes
    | \psi \rangle
    \nonumber \\
    \longmapsto
            | 0 \rangle \otimes | 0 \rangle \otimes
            2 s_{p j, p' j'}^+
            | \psi \rangle
    \nonumber \\
            +
            | 0 \rangle \otimes | 1 \rangle \otimes
            2 \widetilde{s}_{p j, p' j'}^+
            | \psi \rangle
    \nonumber \\
            +
            | 1 \rangle \otimes | 0 \rangle \otimes
            2 s_{p j, p' j'}^-
            | \psi \rangle
    \nonumber \\
            +
            | 1 \rangle \otimes | 1 \rangle \otimes
            2 \widetilde{s}_{p j, p' j'}^-
            | \psi \rangle
    \nonumber \\
    \equiv
        | \Phi_{p j, p' j'} \rangle
    .
\end{gather}
The projective measurement on $| q^{\mathrm{A}}_1 \rangle$ and $| q^{\mathrm{A}}_0 \rangle$ is represented by the four operators
$\mathcal{P}_{q q'} = | q \rangle \langle q | \otimes | q' \rangle \langle q' | \otimes I \, (q, q' = 0, 1)$.
The two outcomes among the possible four are of our interest,
immediately after which the whole system collapses as follows:
\begin{gather}
    | \Phi_{p j, p' j'} \rangle
    \overset{| 0 \rangle \otimes | 0 \rangle \, \mathrm{observed}}{\longmapsto}
        | 0 \rangle
        \otimes
        | 0 \rangle
        \otimes
        \frac{2 s_{p j, p' j'}^+}{\sqrt{\mathbb{P}_{p j, p' j'}^+} }
        | \psi \rangle
    \nonumber \\
    \mathrm{prob.}
    \,
        \left\| 2 s_{p j, p' j'}^+ | \psi \rangle \right\|^2
    \equiv
        \mathbb{P}_{p j, p' j'}^+
    ,
    \label{state_trans_spin_spin_off_diag_0_0}
    \\
    | \Phi_{p j, p' j'} \rangle
    \overset{| 1 \rangle \otimes | 0 \rangle \, \mathrm{observed}}{\longmapsto}
        | 1 \rangle
        \otimes
        | 0 \rangle
        \otimes
        \frac{2 s_{p j, p' j'}^-}{\sqrt{\mathbb{P}_{p j, p' j'}^-} }
        | \psi \rangle
    \nonumber \\
    \mathrm{prob.}
    \,
        \left\| 2 s_{p j, p' j'}^- | \psi \rangle \right\|^2
    \equiv
        \mathbb{P}_{p j, p' j'}^-
    .
    \label{state_trans_spin_spin_off_diag_1_0}
\end{gather}

\subsubsection{Transition matrices via statistical sampling}

We can get the transition matrices $S_\lambda$ via statistical sampling similarly to the charge-charge ones.
If we input $| \Psi_{\mathrm{gs}} \rangle$ to the diagonal circuit $\mathcal{C}_{p j}$ in Fig. \ref{circuit_prep_el_spin_x_y} followed by a measurement and QPE for $\mathcal{H}$,
the energy eigenvalue $E_{\lambda}$ will be obtained with a probability $4 S_{\lambda p j, p j}$.
[See eqs. (\ref{def_transition_mat_for_spin}) and (\ref{state_trans_spin_spin_diag})]
If we use the off-diagonal circuit $\mathcal{C}_{p j, p' j'}$ in Fig. \ref{circuit_prep_el_spin_x_y_aux_using_U},
on the other hand,
the energy eigenvalue $E_{\lambda}$ will be obtained with
probabilities $4 T_{\lambda p j, p' j'}^\pm$ depending on the measurement outcome.
[See eqs. (\ref{state_trans_spin_spin_off_diag_0_0}) and (\ref{state_trans_spin_spin_off_diag_1_0})]
The off-diagonal components of transition matrices are then calculated from eq. (\ref{transition_spin_resp_off_diag_alpha_half}).

\subsection{Spin-charge responses}
\label{subsec:spin_charge_contr}

Having found ways to determine the charge-charge and spin-spin contributions,
let us consider the determination of the spin-charge transition matrices $M_\lambda$.

\subsubsection{Circuits for off-diagonal components}

For combinations $(p, j)$ and $(p', \sigma')$ of spatial orbitals and spin components $j = x, y$ with $\sigma' = \alpha, \beta$,
we define the two non-Hermitian auxiliary operators
\begin{gather}
    v_{p j, p' \sigma'}^\pm
    \equiv
        s_{p j}
        +
        e^{\pm i \pi/4}
        \frac{n_{p' \sigma'}}{2}
    .
    \label{def_opr_spin_occ_aux}
\end{gather}
Unnormalized auxiliary states
$
| \Psi_{p j, p' \sigma'}^\pm \rangle
\equiv
v_{p j, p' \sigma'}^\pm
| \Psi_{\mathrm{gs}} \rangle
$
can have overlaps
$
T_{\lambda p j, p' \sigma'}^\pm
\equiv
| \langle \Psi_\lambda | \Psi_{p j, p' \sigma'}^\pm \rangle |^2
$
with the energy eigenstates,
from which the spin-charge transition matrix elements in eq. (\ref{def_transition_mat_for_spin_occ}) can be calculated as
\begin{gather}
    M_{\lambda p j, p' \sigma'}
    =
        e^{-i \pi/4}
        T_{\lambda p j, p' \sigma'}^+
        +
        e^{i \pi/4}
        T_{\lambda p j, p' \sigma'}^-
    \nonumber\\
        -
        \sqrt{2}
        S_{\lambda p j, p j}
        -
        \frac{N_{\lambda p' \sigma', p' \sigma'}}{2 \sqrt{2}}
        .
    \label{transition_mat_spin_charge_from_T}
\end{gather}

\begin{figure}
\centering
\mbox{ 
\Qcircuit @C=0.6em @R=0.6em { 
    \lstick{| q^{\mathrm{A}}_2 = 0 \rangle} & \gate{H} & \ctrlo{2}                          & \ctrl{1}                                    & \gate{Z \left(\pm \frac{\pi}{4} \right)} & \ctrl{1} & \gate{H} & \meter & \cw \\
    \lstick{| q^{\mathrm{A}}_1 = 0 \rangle} & \gate{H} & \qw                                & \multigate{2}{\mathcal{U}^{(p')}_{\sigma'}} & \gate{H} & \gate{X} & \qw & \qw & \qw \\
    \lstick{| q^{\mathrm{A}}_0 = 0 \rangle} & \gate{H} & \multigate{1}{\mathcal{U}^{(p)}_j} & \ghost{\mathcal{U}^{(p')}_{\sigma'}}        & \gate{H} & \qw      & \qw & \meter & \cw \\
    \lstick{| \psi \rangle}                   & {/} \qw  & \ghost{\mathcal{U}^{(p)}_j}      & \ghost{\mathcal{U}^{(p')}_{\sigma'}}        & \qw      & \qw      & \qw & \qw & \rstick{| \widetilde{\psi} \rangle} \qw
}
} 
\caption{
Spin-charge off-diagonal circuit
$\mathcal{C}_{p j, p' \sigma'}^\pm \, (j = x, y \, \mathrm{and} \, \sigma' = \alpha, \beta)$
for probabilistic preparation of
$v_{p j, p' \sigma'}^\pm | \psi \rangle$
and other states
from an arbitrary input state $| \psi \rangle$ and three ancillary qubits.
The partial circuits defined in Figs.
\ref{circuit_prep_el_occ_alpha_beta} and
\ref{circuit_prep_el_spin_x_y} are contained as the controlled subroutines.
} 
\label{circuit_prep_el_xy_ab_aux_using_U} 
\end{figure}
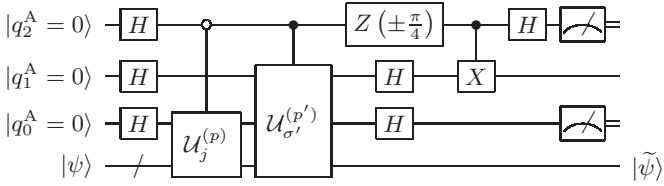

We construct a circuit $\mathcal{C}_{p j, p' \sigma'}^\pm$ equipped with three ancillary qubits by using the controlled operations of the partial circuits $\mathcal{U}^{(p)}_{j}$ in Fig. \ref{circuit_prep_el_spin_x_y} and
$\mathcal{U}^{(p')}_{\sigma'}$ in Fig. \ref{circuit_prep_el_occ_alpha_beta},
as depicted in Fig. \ref{circuit_prep_el_xy_ab_aux_using_U}. 
The whole system consists of the ancillae and an arbitrary input register $| \psi \rangle$.
Its state changes by undergoing the circuit as
\begin{gather}
    | q^{\mathrm{A}}_2 = 0 \rangle
    \otimes
    | q^{\mathrm{A}}_1 = 0 \rangle
    \otimes
    | q^{\mathrm{A}}_0 = 0 \rangle
    \otimes
    | \psi \rangle
    \nonumber \\
    \longmapsto
        | 0 \rangle \otimes | 0 \rangle \otimes | 0 \rangle \otimes
        v_{p j, p' \sigma'}^\pm
        | \psi \rangle
    \nonumber \\
        +
        \left(
            | 0 \rangle \otimes | 0 \rangle \otimes | 1 \rangle
            +
            | 1 \rangle \otimes | 0 \rangle \otimes | 1 \rangle
        \right)
        \otimes
        \widetilde{s}_{p j} | \psi \rangle
    \nonumber \\
        +
        \left(
            | 0 \rangle \otimes | 1 \rangle \otimes | 1 \rangle
            -
            | 1 \rangle \otimes | 1 \rangle \otimes | 1 \rangle
        \right)
        \otimes
        \frac{e^{\pm i \pi/4} \widetilde{n}_{p' \sigma'}}{2} 
        | \psi \rangle
    \nonumber \\
        +
        | 1 \rangle \otimes | 0 \rangle \otimes | 0 \rangle \otimes
        (s_{p j} - e^{\pm i \pi /4} n_{p' \sigma'} )
        | \psi \rangle
    \nonumber \\
    \equiv
        | \Phi_{p j, p' \sigma'}^\pm \rangle
    .
\end{gather}
The projective measurement on $| q^{\mathrm{A}}_2 \rangle$ and $| q^{\mathrm{A}}_0 \rangle$ is represented by the four operators
$\mathcal{P}_{q q'} = | q \rangle \langle q | \otimes I \otimes | q' \rangle \langle q' | \otimes I \, (q, q' = 0, 1)$.
Only one of the four possible outcomes is of our interest,
immediately after which the whole system collapses as follows:
\begin{gather}
    | \Phi_{p j, p' \sigma'}^\pm \rangle
    \overset{| 0 \rangle \otimes | 0 \rangle \, \mathrm{observed}}{\longmapsto}
        | 0 \rangle
        \otimes
        | 0 \rangle
        \otimes
        | 0 \rangle
        \otimes
        \frac{v_{p j, p' \sigma'}^\pm }{\sqrt{\mathbb{P}_{p j, p' \sigma'}^\pm}}
        | \psi \rangle
    \nonumber \\
    \mathrm{prob.}
    \,
        \left\| v_{p j, p' \sigma'}^\pm | \psi \rangle \right\|^2
    \equiv
        \mathbb{P}_{p j, p' \sigma'}^\pm
    .
    \label{state_trans_spin_charge_off_diag_0_0}
\end{gather}

\subsubsection{Transition matrices via statistical sampling}

We can get the transition matrices $M_\lambda$ via statistical sampling similarly to the charge-charge and spin-spin ones.
If we input $| \Psi_{\mathrm{gs}} \rangle$ to the off-diagonal circuit $\mathcal{C}_{p j, p' \sigma'}^\pm$ in Fig. \ref{circuit_prep_el_xy_ab_aux_using_U} followed by a measurement and QPE for $\mathcal{H}$,
the energy eigenvalue $E_{\lambda}$ will be obtained with a probability $T_{\lambda p j, p' \sigma'}^\pm$.
[See eq. (\ref{state_trans_spin_charge_off_diag_0_0})]
The off-diagonal components of transition matrices are then calculated from eq. (\ref{transition_mat_spin_charge_from_T}).

We provide the pseudocodes in Appendix \ref{appendix:pseudocodes} for the calculation procedures of response functions explained above.

\subsection{Generic cases}

Here we describe briefly the scheme for the response function involving generic one-body operators given as eq. (\ref{resp_one_body_def_one_body_opr}).

\subsubsection{Circuits for diagonal components}

\begin{figure*}
\centering
\mbox{ 
\Qcircuit @C=0.8em @R=1em { 
                                                                &          & \mathcal{U}^{(m_1 m_2)} &            &            &            &          &        & \\        
    \lstick{ | q^{\mathrm{A}}_1 = 0 \rangle} & \gate{H} & \ctrlo{1}  & \ctrlo{1}  & \ctrl{1}   & \ctrl{1}   & \gate{H} & \meter & \cw \\
    \lstick{                    | q^{\mathrm{A}}_0 = 0 \rangle} & \gate{H} & \ctrlo{1}  & \ctrl{1}   & \ctrlo{1}  & \ctrl{1}   & \gate{H} & \meter & \cw \\
    \lstick{ | \psi \rangle}                   & {/} \qw  & \gate{U^{(m_1 m_2)}_{0 0}} & \gate{U^{(m_1 m_2)}_{0 1}} & \gate{U^{(m_1 m_2)}_{1 0}} & \gate{U^{(m_1 m_2)}_{1 1}} & \qw      & \qw    & \rstick{| \widetilde{\psi} \rangle} \qw \gategroup{2}{3}{4}{6}{1em}{--}
} 
} 
\vspace{0.5cm}
\caption{
Generic diagonal circuit
$\mathcal{C}_{m_1 m_2}$
for probabilistic preparation of
$a_{m_1}^\dagger a_{m_2} | \psi \rangle$ and other states
from an arbitrary input state $| \psi \rangle$ and
two ancillary qubits.
We define the partial circuit $\mathcal{U}^{(m_1 m_2)}$ by enclosing it with dashed lines.
} 
\label{circuit_prep_one_body_diag} 
\end{figure*}
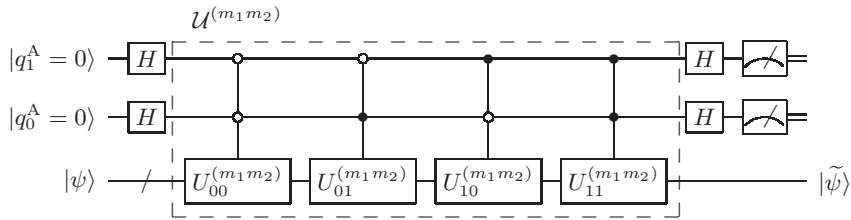

For a combination of spin orbitals $m_1$ and $m_2$,
we define the four unitary operators
$
U^{(m_1 m_2)}_{\kappa \kappa'}
\equiv
U_{\kappa m_1} U_{\kappa' m_2} \, (\kappa, \kappa' = 0, 1)
.
$
We construct a circuit $\mathcal{C}_{m_1 m_2}$ equipped with two ancillary qubits by implementing the controlled operations
for an arbitrary input register $| \psi \rangle$,
as depicted in Fig. \ref{circuit_prep_one_body_diag}.
It is easily confirmed that the whole state changes by undergoing the circuit as
\begin{gather}
    | q^{\mathrm{A}}_1 = 0 \rangle
    \otimes
    | q^{\mathrm{A}}_0 = 0 \rangle
    \otimes
    | \psi \rangle
    \nonumber \\
    \longmapsto
        | 1 \rangle \otimes
        | 0 \rangle \otimes
        a_{m_1}^\dagger a_{m_2} | \psi \rangle
        +
        \mathrm{(other \, terms)}
    \label{resp_one_body_state_transition_diag}
\end{gather}
When the target ground state is input,
the projective measurement on the two ancillae leads to the probabilistic preparation of
$a_{m_1}^\dagger a_{m_2} | \Psi_{\mathrm{gs}} \rangle$,
for which the subsequent QPE process gives 
the ``diagonal'' matrix element
$B_{\lambda m_2 m_1, m_1 m_2}$
[See eq. (\ref{resp_one_body_def_transition_mat})]
via statistical sampling.

The charge-charge diagonal circuit $\mathcal{C}_{p \sigma}$ in Fig. \ref{circuit_prep_el_occ_alpha_beta} is a special case of the generic circuit described here.

\subsubsection{Circuits for off-diagonal components}

For mutually different combinations $(m_1, m_2)$ and $(m_3, m_4)$ of spin orbitals,
we define the two non-Hermitian auxiliary operators
\begin{gather}
    f_{m_1 m_2, m_3 m_4}^\pm
    \equiv
        \frac{
            a_{m_1}^\dagger
            a_{m_2}
            \pm
            e^{i \pi/4}
            a_{m_3}^\dagger
            a_{m_4}
        }{2}
        .
    \label{resp_one_body_def_opr_aux}
\end{gather}
Unnormalized auxiliary states
$
| \Psi_{m_1 m_2, m_3 m_4}^\pm \rangle
\equiv
f_{m_1 m_2, m_3 m_4}^\pm
| \Psi_{\mathrm{gs}} \rangle
$
can have overlaps
$
T_{\lambda m_1 m_2, m_3 m_4}^\pm
\equiv
| \langle \Psi_{\lambda} | \Psi_{m_1 m_2, m_3 m_4}^\pm \rangle |^2
$
with the energy eigenstates,
from which the ``off-diagonal'' matrix element is calculated as
\begin{gather}
    B_{\lambda m_1 m_2, m_3 m_4}
    =   
        e^{-i \pi/4}
        (
            T_{\lambda m_2 m_1, m_3 m_4}^+
            -
            T_{\lambda m_2 m_1, m_3 m_4}^-
        )
    \nonumber \\
        +        
        e^{i \pi/4}
        (
            T_{\lambda m_3 m_4, m_2 m_1}^+
            -
            T_{\lambda m_3 m_4, m_2 m_1}^-
        )
        .
    \label{transition_one_body_off_diag}
\end{gather}

\begin{figure} 
\centering
\mbox{ 
\Qcircuit @C=0.6em @R=0.6em { 
    \lstick{| q^{\mathrm{A}}_2 = 0 \rangle} & \gate{H} & \ctrlo{1}                                  & \ctrl{1} & \gate{Z \left( \frac{\pi}{4} \right)}                    & \gate{H} & \meter & \cw \\
    \lstick{                    | q^{\mathrm{A}}_1 = 0 \rangle} & \gate{H} & \multigate{2}{\mathcal{U}^{(m_2 m_1)}}  & \multigate{2}{\mathcal{U}^{(m_3 m_4)}} & \qw              & \gate{H} & \meter & \cw \\
    \lstick{                    | q^{\mathrm{A}}_0 = 0 \rangle} & \gate{H} & \ghost{\mathcal{U}^{(m_2 m_1)}}         & \ghost{\mathcal{U}^{(m_3 m_4)}} & \qw                    & \gate{H} & \meter & \cw \\
    \lstick{| \psi \rangle}                   & {/} \qw  & \ghost{\mathcal{U}^{(m_2 m_1)}}         & \ghost{\mathcal{U}^{(m_3 m_4)}} & \qw                    & \qw & \qw & \rstick{| \widetilde{\psi} \rangle} \qw
} 
} 
\vspace{0.5cm}
\caption{
Generic off-diagonal circuit
$\mathcal{C}_{m_2 m_1, m_3 m_4}$
for probabilistic preparation of
$f_{m_2 m_1, m_3 m_4}^\pm | \psi \rangle$
and other states from an arbitrary input state $| \psi \rangle$ and three ancillary qubits.
The partial circuits defined in Fig. \ref{circuit_prep_one_body_diag} are contained as the controlled subroutines.
} 
\label{circuit_prep_one_body_off_diag} 
\end{figure}
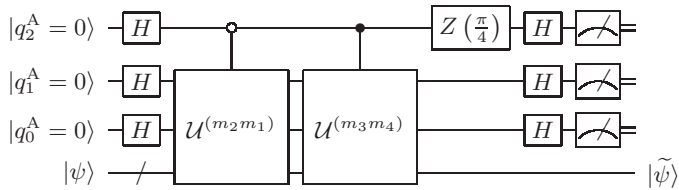

We construct a circuit $\mathcal{C}_{m_2 m_1, m_3 m_4}$ equipped with three ancillary qubits by using the controlled operations of the partial circuits 
defined in Fig. \ref{circuit_prep_one_body_diag},
as depicted in Fig. \ref{circuit_prep_one_body_off_diag}, 
for an arbitrary input register $| \psi \rangle$.
It is easily confirmed that the whole state changes by undergoing the circuit as
\begin{gather}
        | q^{\mathrm{A}}_2 = 0 \rangle
        \otimes
        | q^{\mathrm{A}}_1 = 0 \rangle
        \otimes
        | q^{\mathrm{A}}_0 = 0 \rangle
    \otimes
        | \psi \rangle
    \nonumber \\
    \longmapsto
        | 0 \rangle \otimes
        | 1 \rangle \otimes
        | 0 \rangle \otimes
        f_{m_2 m_1, m_3 m_4}^+ | \psi \rangle
    \nonumber \\
        +
        | 1 \rangle \otimes
        | 1 \rangle \otimes
        | 0 \rangle \otimes
        f_{m_2 m_1, m_3 m_4}^- | \psi \rangle
        +
        \mathrm{(other \, terms)}
        .
    \label{resp_one_body_state_transition_off_diag}
\end{gather}
When the target ground state is input,
the projective measurement on the three ancillae leads to the probabilistic preparation of
$| \Psi_{m_2 m_1, m_3 m_4}^\pm \rangle$.
This circuit and the similarly constructed
$\mathcal{C}_{m_3 m_4, m_2 m_1}$
allows one to calculate 
$B_{\lambda m_1 m_2, m_3 m_4}$
from eq. (\ref{transition_one_body_off_diag}) via statistical sampling.
After obtaining the necessary matrix elements,
they are put into eq. (\ref{resp_one_body_sum_of_transition_mats}) to calculate the Lehmann summation.

The charge-charge off-diagonal circuit $\mathcal{C}_{p \sigma, p' \sigma'}$ in Fig. \ref{circuit_prep_el_occ_alpha_beta_aux_using_U} is a special case of the generic circuit described here.

\section{Computational details}
\label{sec:computational_details}

We adopted STO-6G basis sets as the Cartesian Gaussian-type basis functions\cite{Helgaker} for all the elements in our quantum chemistry calculations.
The two-electron integrals between the atomic orbitals (AOs) were calculated efficiently.\cite{Libint1}
We first performed restricted Hartree--Fock (RHF) calculations to get the orthonormalized molecular orbitals (MOs) in the target systems and
calculated the two-electron integrals between them,
from which we constructed the second-quantized electronic Hamiltonians.

In the FCI calculations for the large target Hilbert subspaces,
we performed exact diagonalization of the electronic (not in qubit representation)
Hamiltonians by using the Arnoldi method.\cite{bib:ARPACK}
We can take the $z$ axis as the quantization axis for spins without loss of generality since our calculations are nonrelativistic.

We calculated the FCI response functions simply by substituting the necessary quantities into eq. (\ref{resp_func_using_R}).
For the simulations of response functions from statistical sampling,
we generated random numbers according to the matrix elements between the FCI energy eigenstates to mimic the measurements on ancillae and the ideal QPE procedures.
We set $\delta$ in eq. (\ref{resp_func_using_R}) to $0.01$ a.u. throughout the present study.

\section{Results and discussion}
\label{sec:results_and_discussion}

\subsection{C$_2$ molecule}

\begin{figure}
\begin{center}
\includegraphics[width=6cm]{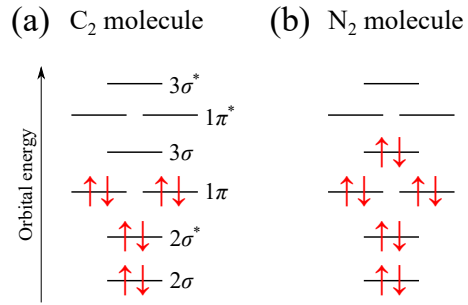}
\end{center}
\caption{
Schematic illustration of RHF orbitals and their electronic occupancies for (a) a C$_2$ molecule and (b) an N$_2$ molecule.
The descriptions beside the energy levels represent the orbital characters.
Those with asterisks are the anti-bonding orbitals. 
The $1 \sigma$ and $1 \sigma^*$ MOs,
coming from the $1 s$ AOs of the constituent atoms,
are not shown.
The $1 \pi$ and $1 \pi^*$ MOs come mainly from the $\pi$ bonding of $2 p$ AOs,
while the $3 \sigma$ and $3 \sigma^*$ MOs from the $\sigma$ bonding of $2 p$ AOs.
The nondegenerate many-electron ground state for each system consists mainly of the same electronic configuration as the RHF one.
}
\label{fig:orb_energies}
\end{figure}

We used the experimental bond length of 1.242 \AA\cite{bib:4867} for a C$_2$ molecule in the RHF calculation and obtained the total energy $E_{\mathrm{RHF}} = -2045.2939$ eV.
This system contains six electrons per spin direction which occupy the lowest six MOs,
as shown in Fig. \ref{fig:orb_energies}(a).
We found via the subsequent FCI calculation with $E_{\mathrm{FCI}} = -2052.6918$ eV that the major electronic configuration in the nondegenerate many-electron ground state,
denoted by $X^1 \Sigma_{\mathrm{g}}^+$ in spectroscopic notation\cite{bib:4868},
is the same as in the RHF solution.
The ground state was found via the exact diagonalization of the Hilbert subspace for $n_\alpha = n_\beta = 6$,
from which we obtained the lowest 2000 among the 44100 energy eigenvalues.

\begin{figure}
\begin{center}
\includegraphics[width=6cm]{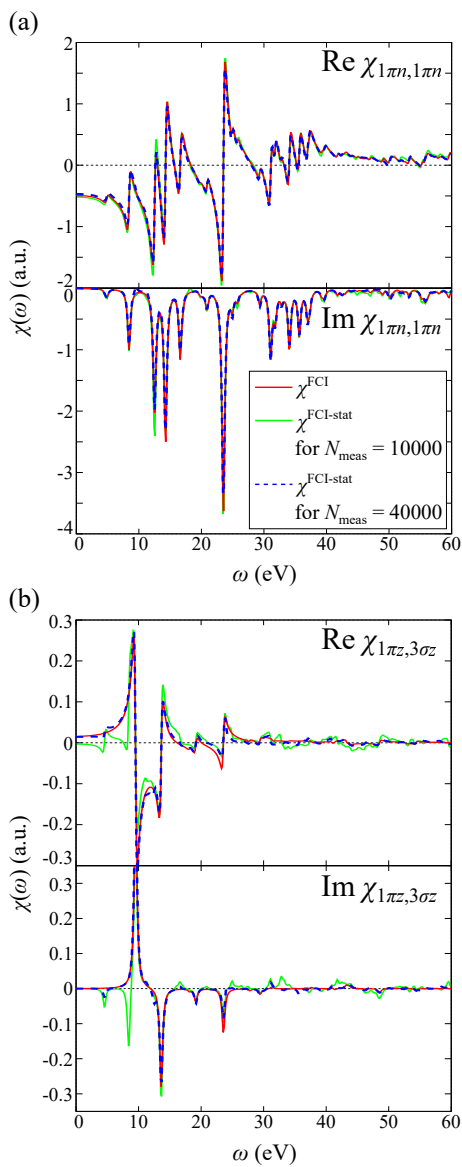}
\end{center}
\caption{
The response function (a) $\chi_{1 \pi n, 1 \pi n}^{\mathrm{FCI}}$ and (b) $\chi_{1 \pi z, 3 \sigma z}^{\mathrm{FCI}}$ exact within the FCI solution for a C$_2$ molecule.
The simulated ones $\chi^{\mathrm{FCI-stat}}$ for for the numbers of measurements $N_{\mathrm{meas}} = 10000$ and $40000$ are also shown.
}
\label{fig:C2_chi}
\end{figure}

We calculated the response functions $\chi^{\mathrm{FCI}}$ exact within the FCI solution,
from which the components $\chi_{1 \pi n, 1 \pi n}^{\mathrm{FCI}}$ and $\chi_{1 \pi z, 3 \sigma z}^{\mathrm{FCI}}$ are plotted in Fig. \ref{fig:C2_chi}.
We also performed simulations of statistical sampling for the construction of response function $\chi^{\mathrm{FCI-stat}}$ based on our scheme and plotted those for $N_{\mathrm{meas}} = 10000$ and $40000$.
It is seen that $\chi_{1 \pi n, 1 \pi n}^{\mathrm{FCI}}$ in Fig. \ref{fig:C2_chi}(a),
which involves only the HOMO,
is well reproduced by
$\chi_{1 \pi n, 1 \pi n}^{\mathrm{FCI-stat}}$
with
$N_{\mathrm{meas}} = 10000$.
On the other hand,
$\chi_{1 \pi z, 3 \sigma z}^{\mathrm{FCI}}$ in Fig. \ref{fig:C2_chi}(b) is not accurately reproduced by
$\chi_{1 \pi n, 1 \pi n}^{\mathrm{FCI-stat}}$
with
$N_{\mathrm{meas}} = 10000$.
These results mean that the response involving a weak excitation channel requires a large number of measurements for its accurate reproduction,
just like the situation for the GFs.\cite{2019arXiv190803902K}

\begin{figure}
\begin{center}
\includegraphics[width=7cm]{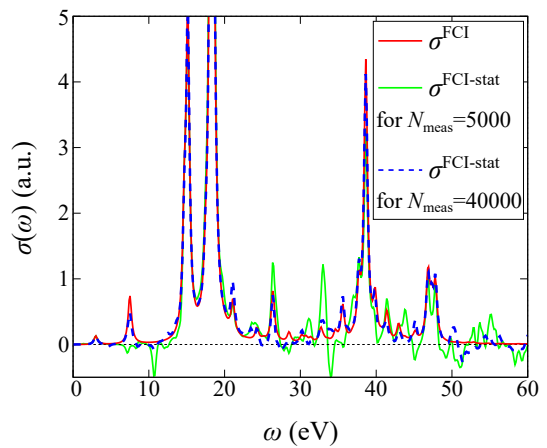}
\end{center}
\caption{
The photoabsorption cross section $\sigma^{\mathrm{FCI}}$ 
exact within the FCI solution for a C$_2$ molecule.
The simulated ones $\sigma^{\mathrm{FCI-stat}}$ for for the numbers of measurements $N_{\mathrm{meas}} = 5000$ and $40000$ are also shown.
}
\label{fig:C2_photo}
\end{figure}

We calculated the photoabsorption cross section $\sigma^{\mathrm{FCI}}$ from the FCI solution,
as shown in Fig. \ref{fig:C2_photo}.
By simulating the process for a generic one-body operator to obtain the matrix elements in eq. (\ref{resp_one_body_def_transition_mat}),
we also calculated the cross section $\sigma^{\mathrm{FCI-stat}}$ for
$N_{\mathrm{meas}} = 5000$ and $40000$ and plotted it in the figure.
It is seen that those from sampling can take negative values
despite the original definition in eq. (\ref{def_photoabs_cross_section}),
which is ensured to be nonnegative.
It is due to our na\"ive implementation of eq. (\ref{transition_one_body_off_diag}) by using random numbers without considering any symmetry.
The major peaks in $\sigma^{\mathrm{FCI}}$ were well reproduced by $\sigma^{\mathrm{FCI-stat}}$ with the smaller $N_{\mathrm{meas}}$,
while the detailed structure between them required the larger $N_{\mathrm{meas}}$ for the accurate reproduction.
Since symmetry-adapted construction of transition matrix elements should reduce the necessary total number of measurements
and physically appropriate results,
such techniques will be useful as well as in VQE calculations.\cite{bib:4999}

\subsection{N$_2$ molecule}

We used the experimental bond length of 1.098 \AA\cite{exp_N2_bond_length} for an N$_2$ molecule in the RHF calculation and obtained the total energy $E_{\mathrm{RHF}} = -2953.5952$ eV.
This system contains seven electrons per spin direction which occupy the lowest seven MOs,
as shown in Fig. \ref{fig:orb_energies}(b).
We found via the subsequent FCI calculation with $E_{\mathrm{FCI}} = -2957.9124$ eV that the major electronic configuration in the nondegenerate many-electron ground state is $X^1 \Sigma_{\mathrm{g}}^+$\cite{bib:4868},
the same as in the RHF solution.
The ground state was found via the exact diagonalization of the Hilbert subspace for $n_\alpha = n_\beta = 7$,
from which we obtained the lowest 2000 among the 14400 energy eigenvalues.

\begin{figure}
\begin{center}
\includegraphics[width=6cm]{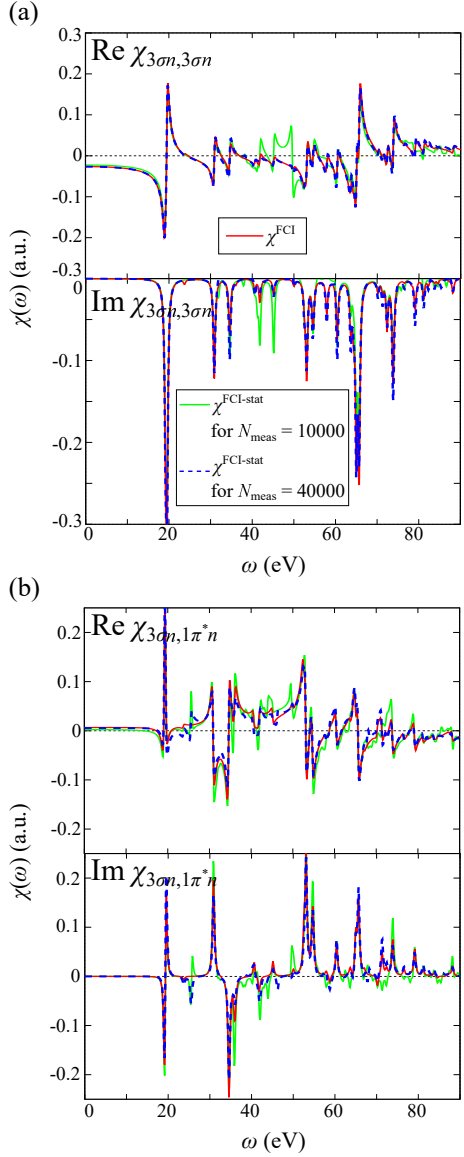}
\end{center}
\caption{
The response function (a) $\chi_{3 \sigma n, 3 \sigma n}^{\mathrm{FCI}}$ and (b) $\chi_{3 \sigma n, 1 \pi^* n}^{\mathrm{FCI}}$ exact within the FCI solution for an N$_2$ molecule.
The simulated ones $\chi^{\mathrm{FCI-stat}}$ for for the numbers of measurements $N_{\mathrm{meas}} = 10000$ and $40000$ are also shown.
}
\label{fig:N2_chi}
\end{figure}

We calculated the response functions from the FCI solution,
from which the components
$\chi_{3 \sigma n, 3 \sigma n}^{\mathrm{FCI}}$ and
$\chi_{3 \sigma n, 1 \pi^* n}^{\mathrm{FCI}}$ are plotted in Fig. \ref{fig:N2_chi}.
We also performed simulations for the construction of $\chi^{\mathrm{FCI-stat}}$ and plotted those for $N_{\mathrm{meas}} = 10000$ and $40000$.
It is seen that $\chi_{3 \sigma n, 3 \sigma n}^{\mathrm{FCI}}$ in Fig. \ref{fig:N2_chi}(a),
which involves only the HOMO,
is not well reproduced by
$\chi_{3 \sigma n, 3 \sigma n}^{\mathrm{FCI-stat}}$
with
$N_{\mathrm{meas}} = 10000$,
in contrast to the C$_2$ molecule case.
Since the strength of $\chi_{3 \sigma n, 1 \pi^* n}^{\mathrm{FCI}}$ in Fig. \ref{fig:N2_chi}(b)
is similar to that of $\chi_{3 \sigma n, 3 \sigma n}^{\mathrm{FCI}}$,
$N_{\mathrm{meas}} = 10000$ is not sufficient for the accurate reproduction of correct values as well.
We found that $\chi_{3 \sigma z, 1 \pi^* z}^{\mathrm{FCI}}$ (not shown) is much weaker and even $N_{\mathrm{meas}} = 40000$ is insufficient for obtaining good $\chi_{3 \sigma z, 1 \pi^* z}^{\mathrm{FCI-stat}}$.

\begin{figure}
\begin{center}
\includegraphics[width=6cm]{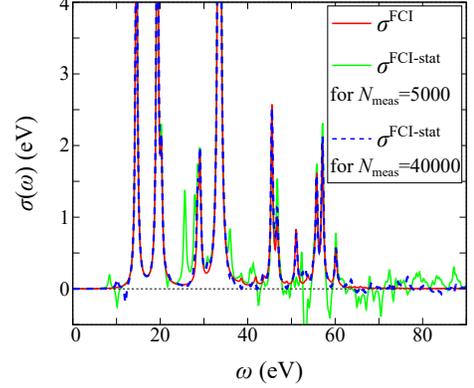}
\end{center}
\caption{
The photoabsorption cross section $\sigma^{\mathrm{FCI}}$ 
exact within the FCI solution for an N$_2$ molecule.
The simulated ones $\sigma^{\mathrm{FCI-stat}}$ for for the numbers of measurements $N_{\mathrm{meas}} = 5000$ and $40000$ are also shown.
}
\label{fig:N2_photo}
\end{figure}

We calculated the photoabsorption cross section $\sigma^{\mathrm{FCI}}$ from the FCI solution,
as shown in Fig. \ref{fig:N2_photo}.
Those from simulations of measurements $\sigma^{\mathrm{FCI-stat}}$ for
$N_{\mathrm{meas}} = 5000$ and $40000$ are also shown in the figure.

\section{Conclusions}
\label{sec:conclusions}

We proposed a scheme for the construction of linear-response functions of an interacting electronic system via QPE and statistical sampling on a quantum computer.
By using the unitary decomposition of electronic operators for avoiding the difficulty due to their nonunitarity, 
we provided the circuits equipped with at most three ancillae for probabilistic preparation of qubit states on which the necessary nonunitary operators have acted.
We performed simulations of such construction of the response functions for C$_2$ and N$_2$ molecules by comparing with the accurate ones based on the FCI calculations.
It was found that the accurate detection of subtle structures coming from the weak poles in the response functions requires a large number of measurements.

Since the unitary decomposition of electronic operators is applicable regardless of the adopted qubit representation,
an electronic state on which an arbitrary product of the creation and annihilation operators has acted can be prepared at least probabilistically.
The approach described in this study thus enables one to access not only the response functions and GFs but also various physical quantities on a quantum computer.
Invention and enrichment of tools for such properties will enhance the practical use of quantum computers for material simulations.

\begin{acknowledgments}

This research was supported by MEXT as Exploratory Challenge on Post-K computer (Frontiers of Basic Science: Challenging the Limits) and Grants-in-Aid for Scientific Research (A) (Grant Numbers 18H03770) from JSPS (Japan Society for the Promotion of Science).

\end{acknowledgments}

\begin{widetext}

\appendix

\section{Proof for a generic probabilistic state preparation}
\label{appendix:proof_prob_stat_prep}

\subsection{Proof}

Here we provide the proof for the validity of circuit $\mathcal{C}_{\mathcal{O}}$ in Fig. \ref{circuit_for_lin_combo_unitaries} for probabilistic state preparation.
$\mathcal{C}_{\mathcal{O}}$ is characterized by the $2^n - 1$ parameters,
$
\theta^{(n)},
\theta^{(n-1)}_0, \theta^{(n-1)}_1,
\theta^{(n-2)}_{00}, \theta^{(n-2)}_{01},
\theta^{(n-2)}_{10}, \theta^{(n-2)}_{11},
\dots,
\theta^{(1)}_{1 \cdots 1}
.
$
For an arbitrary input state $|\psi \rangle$
and the linear combination $\mathcal{O}$ of unitaries in eq. (\ref{def_lin_combo_unitaries}),
$\mathcal{O} | \psi \rangle$ up to a normalization constant
can be prepared by setting the parameters to appropriate values,
as demonstrated below.

We use the notation
$
| \Phi_n \rangle
\equiv
| 0 \rangle^{\otimes n}
\otimes
| \psi \rangle
$,
$H_n \equiv H^{\otimes n} \otimes I$,
and
$
R_y (-2 \theta) | q \rangle
\equiv
| q, \theta \rangle
$
for $q = 0, 1$.
The action of circuit on the initial state $| \Phi_n \rangle$ can be tracked by referring to the recursive definition in Fig. \ref{part_circ_for_lin_combo_unitaries}.
Specifically, we find
\begin{gather}
    \mathcal{C}^{(n)}
    H_n
    | \Phi_n \rangle
    \nonumber \\
    =
        \frac{ | 0, \theta^{(n)} \rangle}{\sqrt{2}}
        \otimes
        \mathcal{C}^{(n-1)}_0
        H_{n - 1}
        | \Phi_{n - 1} \rangle
        +
        \frac{ | 1, \theta^{(n)} \rangle}{\sqrt{2}}
        \otimes
        \mathcal{C}^{(n-1)}_1
        H_{n - 1}
        | \Phi_{n - 1} \rangle
    \nonumber \\
    =
        \frac{ | 0, \theta^{(n)} \rangle}{\sqrt{2}}
        \otimes
        \Bigg[
            \frac{ | 0, \theta^{(n-1)}_0 \rangle}{\sqrt{2}}
            \otimes
            \mathcal{C}^{(n-2)}_{00}
            H_{n - 2}
            | \Phi_{n - 2} \rangle
            +
            \frac{ | 1, \theta^{(n-1)}_0 \rangle}{\sqrt{2}}
            \otimes
            \mathcal{C}^{(n-2)}_{01}
            H_{n - 2}
            | \Phi_{n - 2} \rangle
        \Bigg]
    \nonumber \\
        +
        \frac{ | 1, \theta^{(n)} \rangle}{\sqrt{2}}
        \otimes
        \Bigg[
            \frac{ | 0, \theta^{(n-1)}_1 \rangle}{\sqrt{2}}
            \otimes
            \mathcal{C}^{(n-2)}_{10}
            H_{n - 2}
            | \Phi_{n - 2} \rangle
            +
            \frac{ | 1, \theta^{(n-1)}_1 \rangle}{\sqrt{2}}
            \otimes
            \mathcal{C}^{(n-2)}_{11}
            H_{n - 2}
            | \Phi_{n - 2} \rangle
        \Bigg]
    =
        \cdots
    \nonumber \\
    =
        \frac{1}{2^{n/2}}
        | 0, \theta^{(n)} \rangle
        \otimes
        | 0, \theta^{(n-1)}_0 \rangle
        \otimes
        | 0, \theta^{(n-2)}_{00} \rangle
        \otimes
        \cdots
        \otimes
        | 0, \theta^{(1)}_{0 \cdots 0} \rangle
        \otimes
        U_{0 \cdots 0}
        | \psi \rangle
        +
        \cdots
    \nonumber \\
        +
        \frac{1}{2^{n/2}}
        | 1, \theta^{(n)} \rangle
        \otimes
        | 1, \theta^{(n-1)}_1 \rangle
        \otimes
        | 1, \theta^{(n-2)}_{11} \rangle
        \otimes
        \cdots
        \otimes
        | 1, \theta^{(1)}_{1 \cdots 1} \rangle
        \otimes
        U_{1 \cdots 1}
        | \psi \rangle
        .
    \label{action_of_circuit_for_generic_prob_stat_prep}
\end{gather}
For each bit string $k$ of length $n$,
$U_k$ appears only once on the right-hand side of eq. (\ref{action_of_circuit_for_generic_prob_stat_prep}).
If the outcome of a projective measurement on the $n$ ancillae is $| 0 \rangle^{\otimes n}$,
the whole state collapses immediately after the measurement as follows:
\begin{gather}
    \mathcal{C}^{(n)}
    H_n
    | \Phi_n \rangle
    \overset{| 0 \rangle^{\otimes n} \, \mathrm{observed}}{\longmapsto}
        | 0 \rangle^{\otimes n}
        \otimes
        \frac{1}{2^{n/2}}
        \Bigg[
    \nonumber \\
        \cos \theta^{(n)}
        \cos \theta^{(n-1)}_0
        \cos \theta^{(n-2)}_{00}
        \cdots
        \cos \theta^{(2)}_{0 \cdots 0}
        \cos \theta^{(1)}_{0 \cdots 00}
        U_{0 \cdots 000}
    \nonumber \\
        +
        \cos \theta^{(n)}
        \cos \theta^{(n-1)}_0
        \cos \theta^{(n-2)}_{00}
        \cdots
        \cos \theta^{(2)}_{0 \cdots 0}
        \sin \theta^{(1)}_{0 \cdots 00}
        U_{0 \cdots 001}
    \nonumber \\
        \vdots
    \nonumber \\
        +
        \sin \theta^{(n)}
        \sin \theta^{(n-1)}_1
        \sin \theta^{(n-2)}_{11}
        \cdots
        \sin \theta^{(2)}_{1 \cdots 1}
        \cos \theta^{(1)}_{1 \cdots 11}
        U_{1 \cdots 110}
    \nonumber \\
        +
        \sin \theta^{(n)}
        \sin \theta^{(n-1)}_1
        \sin \theta^{(n-2)}_{11}
        \cdots
        \sin \theta^{(2)}_{1 \cdots 1}
        \sin \theta^{(1)}_{1 \cdots 11}
        U_{1 \cdots 111}
        \Bigg]
        | \psi \rangle
\end{gather}
We derived the expression above by using
$
| 0, \theta \rangle
=
\cos \theta | 0 \rangle 
-
\sin \theta | 1 \rangle 
$
and
$
| 1, \theta \rangle
=
\sin \theta | 0 \rangle 
+
\cos \theta | 1 \rangle 
.
$

In order for the resultant state to be proportional to $\mathcal{O} | \psi \rangle$,
we first determine the $2^{n - 1}$ parameters
$\theta^{(1)}_{0 \cdots 00}, \dots, \theta^{(1)}_{1 \cdots 11}$
(each having a bit string of length $n - 1$)
such that
\begin{gather}
    \tan \theta^{(1)}_{0 \cdots 00}
    =
        \frac{c_{0 \cdots 001}}{c_{0 \cdots 000}}
    , \,
    \tan \theta^{(1)}_{0 \cdots 01}
    =
        \frac{c_{0 \cdots 011}}{c_{0 \cdots 010}}
    , \dots,
    \tan \theta^{(1)}_{1 \cdots 11}
    =
        \frac{c_{1 \cdots 111}}{c_{1 \cdots 110}}
    .
    \label{}
\end{gather}
Next we determine the $2^{n - 2}$ parameters
$\theta^{(2)}_{0 \cdots 00}, \dots, \theta^{(2)}_{1 \cdots 11}$
(each having a bit string of length $n - 2$)
such that
\begin{gather}
    \frac{\cos \theta^{(1)}_{0 \cdots 001}}{\cos \theta^{(1)}_{0 \cdots 000}}
    \tan \theta^{(2)}_{0 \cdots 00}
    =
        \frac{c_{0 \cdots 0010}}{c_{0 \cdots 0000}}
    , \,
    \frac{\cos \theta^{(1)}_{0 \cdots 011}}{\cos \theta^{(1)}_{0 \cdots 010}}
    \tan \theta^{(2)}_{0 \cdots 01}
    =
        \frac{c_{0 \cdots 0110}}{c_{0 \cdots 0100}}
    , \dots,
    \frac{\cos \theta^{(1)}_{1 \cdots 111}}{\cos \theta^{(1)}_{1 \cdots 110}}
    \tan \theta^{(2)}_{1 \cdots 11}
    =
        \frac{c_{1 \cdots 1110}}{c_{1 \cdots 1100}}
        .
\end{gather}
All the remaining parameters can also be determined in this way,
so that the circuit $\mathcal{C}_{\mathcal{O}}$ allows one to prepare the desired state probabilistically.

\subsection{Example for $n = 2$}

An example for the determination of parameters for
$\mathcal{C}_\mathcal{O}$ with $n = 2$ is provided here.
The circuit is parametrized by 
$\theta^{(1)}_0, \theta^{(1)}_1$, and $\theta^{(2)}$,
as depicted in Fig. \ref{circuit_for_lin_combo_unitaries_n_2}.
The whole state immediately before a measurement is
\begin{gather}
    \mathcal{C}^{(2)}
    H_2
    | \Phi_2 \rangle
    =
        | 0 \rangle^{\otimes 2}
        \otimes
        \frac{1}{2}
        \left[
        \cos \theta^{(2)}
        \cos \theta^{(1)}_{0}
        U_{00}
        +
        \cos \theta^{(2)}
        \sin \theta^{(1)}_{0}
        U_{01}
        +
        \sin \theta^{(2)}
        \cos \theta^{(1)}_{1}
        U_{10}
        +
        \sin \theta^{(2)}
        \sin \theta^{(1)}_{1}
        U_{11}
        \right]
        | \psi \rangle
    \nonumber \\
        +
        (\mathrm{terms \, involving \, other \, ancillary \, states})
        .
\end{gather}
The probabilistic state preparation is possible by setting the three parameters in the manner described above.
$\theta^{(1)}_0$ and $\theta^{(1)}_1$ are determined by the conditions
\begin{gather}
    \tan \theta^{(1)}_{0}
    =
        \frac{c_{01}}{c_{00}}
    , \,
    \tan \theta^{(1)}_{1}
    =
        \frac{c_{11}}{c_{10}}
    ,
\end{gather}
from which 
$\theta^{(2)}$ is determined by the condition
\begin{gather}
    \frac{\cos \theta^{(1)}_{1}}{\cos \theta^{(1)}_{0}}
    \tan \theta^{(2)}
    =
        \frac{c_{10}}{c_{00}}
        .
\end{gather}

\begin{figure*}
\centering
\mbox{ 
\Qcircuit @C=.5em @R=.8em {
    \lstick{| q^{\mathrm{A}}_1 = 0 \rangle} & \gate{H} & \ctrlo{1}     & \ctrlo{1}     & \ctrlo{1}                                  & \ctrl{1}      & \ctrl{1}      & \ctrl {1}                                  & \gate{R_y \left(-2 \theta^{(2)} \right)} & \meter & \cw\\
    \lstick{| q^{\mathrm{A}}_0 = 0 \rangle} & \gate{H} & \ctrlo{1}     & \ctrl{1}      & \gate{R_y \left(-2 \theta^{(1)}_0 \right)} & \ctrlo{1}     & \ctrl{1}      & \gate{R_y \left(-2 \theta^{(1)}_1 \right)} & \qw                                   & \meter & \cw \\
    \lstick{| \psi    \rangle}              & {/} \qw  & \gate{U_{00}} & \gate{U_{01}} & \qw                                        & \gate{U_{10}} & \gate{U_{11}} & \qw                                        & \qw                                   & \qw      & \qw
} 
} 
\caption{
Circuit $\mathcal{C}_{\mathcal{O}}$
for $n = 2$ as a special case of that in Fig. \ref{circuit_for_lin_combo_unitaries}.
} 
\label{circuit_for_lin_combo_unitaries_n_2} 
\end{figure*}
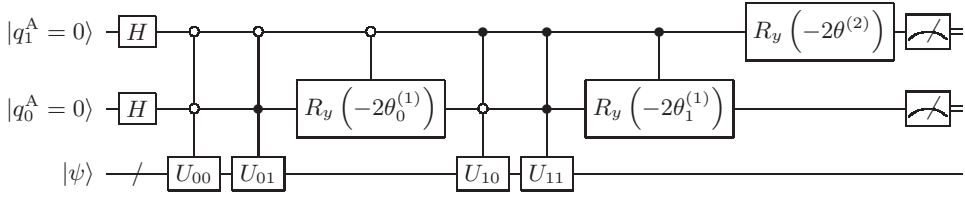

\section{Pseudocodes}
\label{appendix:pseudocodes}

Here we provide the pseudocodes for the calculation of response functions proposed in the present study.
We assume that
all the energy eigenvalues of $N$-electron states
have been obtained before entering the main procedure given just below.

\subsection{Main procedure}

Here is the main procedure.
It calls directly or indirectly all the other procedures provided in this Appendix.

\begin{algorithm}[H]
	\caption{Calculation of charge and spin response functions via statistical sampling}
    \label{alg:CalcRespFuncs}
	\begin{algorithmic}[1]
    	\Require 
    	    \Statex {
    	        Hamiltonian $\mathcal{H}$,
    	        number of spatial orbitals $n_{\mathrm{orbs}}$,
    	        $N$-electron ground state $| \Psi_{\mathrm{gs}} \rangle$
    	        with its energy $E^N_{\mathrm{gs}}$,
    	        energy eigenvalues $E^N_\lambda$,
    	        real frequency $\omega$,
    	        small positive constant $\delta$,
    	        number of measurements $N_{\mathrm{meas}}$ for each component
    	        }
		\Ensure
        	\Statex {Response functions $\chi (\omega)$}
		\Function{CalcRespFuncs}{$
            \mathcal{H}, 
            n_{\mathrm{orbs}},
            | \Psi_{\mathrm{gs}} \rangle,
            E^N_{\mathrm{gs}},
		    E^N,
		    \omega,
		    \delta,
		    N_{\mathrm{meas}}
		    $}
    		\For{$\lambda$}
    		    \State
    		    $
    		    d_{\lambda \pm} :=
    		    \pm (\omega + i \delta) - (E^N_\lambda - E_{\mathrm{gs}}^N)
    		    $
        	\EndFor
		    \State $\chi^{\mathrm{tmp}} := 0$
    		\For{$p = 1, \dots, n_{\mathrm{orbs}}$}
            \Comment{Diagonal components}
        		\For{$\sigma = \alpha, \beta$}
                \Comment{Charge-charge contributions}
        		    \State $N_{p \sigma, p \sigma} :=$ \textsc{AmplsChargeDiag}$(\mathcal{H}, | \Psi_{\mathrm{gs}} \rangle, E^N, p, \sigma, N_{\mathrm{meas}})$
            		\For{$\lambda$}
            		    \State
            		    $\switch
            		    \chi^{\mathrm{tmp}}_{p \sigma, p \sigma} +=
            		    N_{\lambda p \sigma, p \sigma}/ d_{\lambda +}
            		    +
            		    N_{\lambda p \sigma, p \sigma}/ d_{\lambda -}
            		    $
            		\EndFor
        		\EndFor
        		\For{$j = x, y$}
            	\Comment{Spin-spin contributions}
        		    \State $S_{p j, p j} :=$ \textsc{AmplsSpinDiag}$(\mathcal{H}, | \Psi_{\mathrm{gs}} \rangle, E^N, p, j, N_{\mathrm{meas}})$
            		\For{$\lambda$}
            		    \State
            		    $\switch
            		    \chi^{\mathrm{tmp}}_{p j, p j} +=
            		    S_{\lambda p j, p j}/ d_{\lambda +}
            		    +
            		    S_{\lambda p j, p j}/ d_{\lambda -}
            		    $
            		\EndFor
        		\EndFor
    		\EndFor
    		\For{$p, p' = 1, \dots, n_{\mathrm{orbs}} \, (p \geqq p')$}
            \Comment{Off-Diagonal components}
        		\For{$\sigma, \sigma' = \alpha, \beta$}
                \Comment{Charge-charge contributions}
        		    \If{$p > p'$ {\bf or} ($p == p'$ {\bf and} $\sigma == \beta$ {\bf and} $\sigma' == \alpha$)}
                	    \State $N_{p \sigma, p' \sigma'} :=$ \textsc{AmplsChargeOffDiag}$(\mathcal{H}, | \Psi_{\mathrm{gs}} \rangle, E^N, p, p', \sigma, \sigma', N_{\mathrm{meas}})$
                    	\For{$\lambda$}
                	        \State
                	        $\switch
                    		\chi^{\mathrm{tmp}}_{p \sigma, p' \sigma'} +=
                    		N_{\lambda p \sigma, p' \sigma'}/ d_{\lambda +}
                	       	+
                	        N_{\lambda p \sigma, p' \sigma'}^*/ d_{\lambda -}
                	        , \,
                    		\chi^{\mathrm{tmp}}_{p' \sigma', p \sigma} +=
                    		N_{\lambda p \sigma, p' \sigma'}^*/ d_{\lambda +}
                	       	+
                	        N_{\lambda p \sigma, p' \sigma'}/ d_{\lambda -}
                    		$
                		\EndFor
	                \EndIf
                \EndFor
        		\For{$j, j' = x, y$}
                \Comment{Spin-spin contributions}
        		    \If{$p > p'$ {\bf or} ($p == p'$ {\bf and} $j == y$ {\bf and} $j' == x$)}
                	    \State $S_{p j, p' j'} :=$ \textsc{AmplsSpinOffDiag}$(\mathcal{H}, | \Psi_{\mathrm{gs}} \rangle, E^N, p, p', j, j', N_{\mathrm{meas}})$
                    	\For{$\lambda$}
                	        \State
                	        $\switch
                    		\chi^{\mathrm{tmp}}_{p j, p' j'} +=
                    		S_{\lambda p j, p' j'}/ d_{\lambda +}
                	       	+
                	        S_{\lambda p j, p' j'}^*/ d_{\lambda -}
                	        , \,
                    		\chi^{\mathrm{tmp}}_{p' j', p j} +=
                    		S_{\lambda p j, p' j'}^*/ d_{\lambda +}
                	       	+
                	        S_{\lambda p j, p' j'}/ d_{\lambda -}
                    		$
                		\EndFor
	                \EndIf
                \EndFor
        		\For{$j = x, y, \sigma' = \alpha, \beta$}
                \Comment{Spin-charge contributions}
            	        \State $M_{p j, p' \sigma'} :=$
            	        \textsc{AmplsSpinChargeOffDiag}$(
            	        \mathcal{H},
            	        | \Psi_{\mathrm{gs}} \rangle,
            	        E^N,
            	        p, p', j, \sigma',
            	        S_{p j, p j},
            	        N_{p' \sigma', p' \sigma'}, N_{\mathrm{meas}})$
                    	\For{$\lambda$}
                	        \State
                	        $\switch
                    		\chi^{\mathrm{tmp}}_{p j, p' \sigma'} +=
                    		M_{\lambda p j, p' \sigma'}/ d_{\lambda +}
            	           	+
            	            M_{\lambda p j, p' \sigma'}^*/ d_{\lambda -}
                	        , \,
                    		\chi^{\mathrm{tmp}}_{p' \sigma', p j} +=
                    		M_{\lambda p j, p' \sigma'}^*/ d_{\lambda +}
                	       	+
            	            M_{\lambda p j, p' \sigma'}/ d_{\lambda -}
                	    	$
                		\EndFor
                \EndFor
            \EndFor
    		\For{$p, p' = 1, \dots, n_{\mathrm{orbs}}$}
    		\Comment{Representation for $(n, x, y, z)$ from $(\alpha, \beta, x, y)$}
    	        \State
    	        $
    	        \chi_{p n, p' n} :=
    	            \chi^{\mathrm{tmp}}_{p \alpha, p' \alpha} +
    	            \chi^{\mathrm{tmp}}_{p \alpha, p' \beta} +
    	            \chi^{\mathrm{tmp}}_{p \beta, p' \alpha} +
    	            \chi^{\mathrm{tmp}}_{p \beta, p' \beta}
    	        , \,         
    	        \chi_{p n, p' z} :=
    	            (\chi^{\mathrm{tmp}}_{p \alpha, p' \alpha} -
    	            \chi^{\mathrm{tmp}}_{p \alpha, p' \beta} +
    	            \chi^{\mathrm{tmp}}_{p \beta, p' \alpha} -
    	            \chi^{\mathrm{tmp}}_{p \beta, p' \beta})/2
    	        $
    	        \State
    	        $
    	        \chi_{p z, p' n} :=
    	            (\chi^{\mathrm{tmp}}_{p \alpha, p' \alpha} +
    	            \chi^{\mathrm{tmp}}_{p \alpha, p' \beta} -
    	            \chi^{\mathrm{tmp}}_{p \beta, p' \alpha} -
    	            \chi^{\mathrm{tmp}}_{p \beta, p' \beta})/2
    	        , \,
    	        \chi_{p z, p' z} :=
    	            (\chi^{\mathrm{tmp}}_{p \alpha, p' \alpha} -
    	            \chi^{\mathrm{tmp}}_{p \alpha, p' \beta} -
    	            \chi^{\mathrm{tmp}}_{p \beta, p' \alpha} +
    	            \chi^{\mathrm{tmp}}_{p \beta, p' \beta})/4
                $
        		\For{$j = x, y$}
        	        \State
        	        $
        	        \chi_{p n, p' j} :=
        	        \chi^{\mathrm{tmp}}_{p \alpha, p' j}
        	        +
        	        \chi^{\mathrm{tmp}}_{p \beta, p' j}
        	        , \,
        	        \chi_{p z, p' j} :=
        	        (
        	        \chi^{\mathrm{tmp}}_{p \alpha, p' j}
        	        -
        	        \chi^{\mathrm{tmp}}_{p \beta, p' j}
        	        )/2
        	        $
        	        \State
        	        $
        	        \chi_{p j, p' n} :=
        	        \chi^{\mathrm{tmp}}_{p j, p' \alpha}
        	        +
        	        \chi^{\mathrm{tmp}}_{p j, p' \beta}
        	        , \,
        	        \chi_{p j, p' z} :=
        	        (
        	        \chi^{\mathrm{tmp}}_{p j, p' \alpha}
        	        -
        	        \chi^{\mathrm{tmp}}_{p j, p' \beta}
        	        )/2
        	        $
        		\EndFor
        		\For{$j, j' = x, y$}
        	        \State
        	        $\chi_{p j, p' j'} := \chi^{\mathrm{tmp}}_{p j, p' j'}$
        		\EndFor
    		\EndFor
    		\State \Return $\chi$
        \EndFunction
	\end{algorithmic}
\end{algorithm}

\subsection{Charge-charge contributions}

The following procedures calculate the matrix elements in the manner described in Subsection \ref{subsec:chrage_charge_contr}.

\subsubsection{Diagonal components}

\begin{algorithm}[H]
	\caption{Calculation of diagonal components of transition matrices}
    \label{alg:AmplsChargeDiag}
	\begin{algorithmic}[1]
		\Function{AmplsChargeDiag}{$
		    \mathcal{H},
		    | \Psi_{\mathrm{gs}} \rangle,
		    E^N,
		    p,
		    \sigma,
		    N_{\mathrm{meas}}
		    $}
            \State $N_{p \sigma, p \sigma} := 0$
    		\For{$m = 1, \dots, N_{\mathrm{meas}}$}
    		    \State Input $| \Psi_{\mathrm{gs}} \rangle$ to $\mathcal{C}_{p \sigma}$ and measure the ancilla $| q^{\mathrm{A}}_0 \rangle$
    		    \Comment{Circuit in Fig. \ref{circuit_prep_el_occ_alpha_beta}}
    		    \If{$| q_0^{\mathrm{A}} \rangle == | 0 \rangle$}
        		    \State $E :=$ QPE$(| \widetilde{\Psi} \rangle, \mathcal{H})$,
        		        Find $E$ among $\{ E_\lambda^N \}_\lambda$
        		    \Comment{For the register $| \widetilde{\Psi} \rangle$ coming out of the circuit}
                    \State $\switch N_{\lambda p \sigma, p \sigma} += 1$
    		    \EndIf
    		\EndFor
            \State $\switch
                N_{p \sigma, p \sigma} *= 1/N_{\mathrm{meas}}
                $
            \State \Return $N_{p \sigma, p \sigma}$
        \EndFunction
	\end{algorithmic}
\end{algorithm}

\subsubsection{Off-diagonal components}

\begin{algorithm}[H]
	\caption{Calculation of off-diagonal components of transition matrices}
    \label{alg:AmplsChargeOffDiag}
	\begin{algorithmic}[1]
		\Function{AmplsChargeOffDiag}{$
		    \mathcal{H},
		    | \Psi_{\mathrm{gs}} \rangle,
		    E^N,
		    p, p', \sigma, \sigma',
		    N_{\mathrm{meas}}
		    $}
		    \State $T^{\pm}_{p \sigma, p' \sigma'} :=$
		        \textsc{AmplsChargeAux}$(
		        \mathcal{H},
		        | \Psi_{\mathrm{gs}} \rangle, E^N,
		        p, p', \sigma, \sigma',
		        N_{\mathrm{meas}})$
		    \State $T^{\pm}_{p' \sigma', p \sigma} :=$
		        \textsc{AmplsChargeAux}$(
		        \mathcal{H},
		        | \Psi_{\mathrm{gs}} \rangle, E^N,
		        p', p, \sigma', \sigma,
		        N_{\mathrm{meas}})$
    		\For{$\lambda$}
                \State $
                    N_{\lambda p \sigma, p' \sigma'}
                    :=
                        e^{-i \pi/4}
                        (
                            T_{\lambda p \sigma, p' \sigma'}^+
                            -
                            T_{\lambda p \sigma, p' \sigma'}^-
                        )
                        +
                        e^{i \pi/4}
                        (
                            T_{\lambda p' \sigma', p \sigma}^+
                            -
                            T_{\lambda p' \sigma', p \sigma}^-
                        )
                $
                \Comment{See eq. (\ref{transition_occ_off_diag_alpha_half})}
            \EndFor
            \State \Return $N_{p \sigma, p' \sigma'}$
        \EndFunction
	\end{algorithmic}
\end{algorithm}

\begin{algorithm}[H]
	\caption{Calculation of transition amplitudes for auxiliary states}
    \label{alg:AmplsChargeAux}
	\begin{algorithmic}[1]
		\Function{AmplsChargeAux}{$
		    \mathcal{H},
		    | \Psi^N_{\mathrm{gs}} \rangle,
		    E^N,
		    p, p', \sigma, \sigma',
		    N_{\mathrm{meas}}
		    $}
            \State $T^{\pm}_{p \sigma, p' \sigma'} := 0$
    		\For{$m = 1, \dots, N_{\mathrm{meas}}$}
    		    \State Input $| \Psi_{\mathrm{gs}} \rangle$ to $\mathcal{C}_{p \sigma, p' \sigma'}$ and measure the ancillae
    		        $| q_2^{\mathrm{A}} \rangle$ and
    		        $| q_1^{\mathrm{A}} \rangle$
    		    \Comment{Circuit in Fig. \ref{circuit_prep_el_occ_alpha_beta_aux_using_U}}
                \If{$
                    | q_2^{\mathrm{A}} \rangle
                    \otimes
                    | q_1^{\mathrm{A}} \rangle
                    ==
                    | 0 \rangle
                    \otimes
                    | 1 \rangle
                    $}
        		    \State $E :=$ QPE$(| \widetilde{\Psi} \rangle, \mathcal{H})$,
        		        Find $E$ among $\{ E_\lambda^N \}_\lambda$
        		    \Comment{For the register $| \widetilde{\Psi} \rangle$ coming out of the circuit}
    		        \State $\switch T^+_{\lambda p \sigma, p' \sigma'} += 1$
                \Else {\, {\bf if} \, $
                    | q_2^{\mathrm{A}} \rangle
                    \otimes
                    | q_1^{\mathrm{A}} \rangle
                    ==
                    | 1 \rangle
                    \otimes
                    | 1 \rangle
                    $ \, {\bf then}}
        		    \State $E :=$ QPE$(| \widetilde{\Psi} \rangle, \mathcal{H})$,
                        Find $E$ among $\{ E_\lambda^N \}_\lambda$
        		    \Comment{For the register $| \widetilde{\Psi} \rangle$ coming out of the circuit}
                    \State $\switch T^-_{\lambda p \sigma, p' \sigma'} += 1$
    		    \EndIf
    		\EndFor
            \State $\switch
                T^{\pm}_{p \sigma, p' \sigma'} *= 1/N_{\mathrm{meas}}$
            \State \Return $T^{\pm}_{p \sigma, p' \sigma'}$
        \EndFunction
	\end{algorithmic}
\end{algorithm}

\subsection{Spin-spin contributions}

The following procedures calculate the matrix elements in the manner described in Subsection \ref{subsec:spin_spin_contr}.

\subsubsection{Diagonal components}

\begin{algorithm}[H]
	\caption{Calculation of diagonal components of transition matrices}
    \label{alg:AmplsSpinDiag}
	\begin{algorithmic}[1]
		\Function{AmplsSpinDiag}{$
		    \mathcal{H},
		    | \Psi_{\mathrm{gs}} \rangle,
		    E^N,
		    p,
		    j,
		    N_{\mathrm{meas}}
		    $}
            \State $S_{p j, p j} := 0$
    		\For{$m = 1, \dots, N_{\mathrm{meas}}$}
    		    \State Input $| \Psi_{\mathrm{gs}} \rangle$ to $\mathcal{C}_{p j}$ and measure the ancilla $| q^{\mathrm{A}} \rangle$
    		    \Comment{Circuit in Fig. \ref{circuit_prep_el_spin_x_y}}
    		    \If{$| q^{\mathrm{A}} \rangle == | 0 \rangle$}
        		    \State $E :=$ QPE$(| \widetilde{\Psi} \rangle, \mathcal{H})$,
            		    Find $E$ among $\{ E_\lambda^N \}_\lambda$
        		    \Comment{For the register $| \widetilde{\Psi} \rangle$ coming out of the circuit}
                    \State $\switch S_{\lambda p j, p j} += 1$
    		    \EndIf
    		\EndFor
            \State $\switch
                S_{p j, p j} *= 1/(4 N_{\mathrm{meas}})
                $
            \State \Return $S_{p j, p j}$
        \EndFunction
	\end{algorithmic}
\end{algorithm}

\subsubsection{Off-diagonal components}

\begin{algorithm}[H]
	\caption{Calculation of off-diagonal components of transition matrices}
    \label{alg:AmplsSpinOffDiag}
	\begin{algorithmic}[1]
		\Function{AmplsSpinOffDiag}{$
		    \mathcal{H},
		    | \Psi_{\mathrm{gs}} \rangle,
		    E^N,
		    p, p', j, j',
		    N_{\mathrm{meas}}
		    $}
		    \State $T^{\pm}_{p j, p' j'} :=$
		        \textsc{AmplsSpinAux}$(
		        \mathcal{H},
		        | \Psi_{\mathrm{gs}} \rangle, E^N,
		        p, p', j, j',
		        N_{\mathrm{meas}})$
		    \State $T^{\pm}_{p' j', p j} :=$
		        \textsc{AmplsSpinAux}$(
		        \mathcal{H},
		        | \Psi_{\mathrm{gs}} \rangle, E^N,
		        p', p, j', j,
		        N_{\mathrm{meas}})$
    		\For{$\lambda$}
                \State $
                    S_{\lambda p j, p' j'}
                    :=
                        e^{-i \pi/4}
                        (
                            T_{\lambda p j, p' j'}^+
                            -
                            T_{\lambda p j, p' j'}^-
                        )
                        +
                        e^{i \pi/4}
                        (
                            T_{\lambda p' j', p j}^+
                            -
                            T_{\lambda p' j', p j}^-
                        )
                $
                \Comment{See eq. (\ref{transition_spin_resp_off_diag_alpha_half})}
            \EndFor
            \State \Return $S_{p j, p' j'}$
        \EndFunction
	\end{algorithmic}
\end{algorithm}

\begin{algorithm}[H]
	\caption{Calculation of transition amplitudes for auxiliary states}
    \label{alg:AmplsSpinAux}
	\begin{algorithmic}[1]
		\Function{AmplsSpinAux}{$
		    \mathcal{H},
		    | \Psi_{\mathrm{gs}} \rangle,
		    E^N,
		    p, p', j, j',
		    N_{\mathrm{meas}}
		    $}
            \State $T^{\pm}_{p j, p' j'} := 0$
    		\For{$m = 1, \dots, N_{\mathrm{meas}}$}
    		    \State Input $| \Psi_{\mathrm{gs}} \rangle$ to $\mathcal{C}_{p j, p' j'}$ and measure the ancillae
    		        $| q_1^{\mathrm{A}} \rangle$ and
    		        $| q_0^{\mathrm{A}} \rangle$
    		    \Comment{Circuit in Fig. \ref{circuit_prep_el_spin_x_y_aux_using_U}}
                \If{$
                    | q_1^{\mathrm{A}} \rangle
                    \otimes
                    | q_0^{\mathrm{A}} \rangle
                    ==
                    | 0 \rangle
                    \otimes
                    | 0 \rangle
                    $}
        		    \State $E :=$ QPE$(| \widetilde{\Psi} \rangle, \mathcal{H})$,
            		    Find $E$ among $\{ E_\lambda^N \}_\lambda$
        		    \Comment{For the register $| \widetilde{\Psi} \rangle$ coming out of the circuit}
    		        \State $\switch T^+_{\lambda p j, p' j'} += 1$
                \Else {\, {\bf if} \, $
                    | q_1^{\mathrm{A}} \rangle
                    \otimes
                    | q_0^{\mathrm{A}} \rangle
                    ==
                    | 1 \rangle
                    \otimes
                    | 0 \rangle
                    $ \, {\bf then}}
        		    \State $E :=$ QPE$(| \widetilde{\Psi} \rangle, \mathcal{H})$,
            		    Find $E$ among $\{ E_\lambda^N \}_\lambda$
        		    \Comment{For the register $| \widetilde{\Psi} \rangle$ coming out of the circuit}
                    \State $\switch T^-_{\lambda p j, p' j'} += 1$
    		    \EndIf
    		\EndFor
            \State $\switch
                T^{\pm}_{p j, p' j'} *= 1/(4 N_{\mathrm{meas}})$
            \State \Return $T^{\pm}_{p j, p' j'}$
        \EndFunction
	\end{algorithmic}
\end{algorithm}

\subsection{Spin-charge contributions}

The following procedures calculate the matrix elements in the manner described in Subsection \ref{subsec:spin_charge_contr}.

\begin{algorithm}[H]
	\caption{Calculation of off-diagonal components of transition matrices}
    \label{alg:AmplsSpinChargeOffDiag}
	\begin{algorithmic}[1]
		\Function{AmplsSpinChargeOffDiag}{$
		    \mathcal{H},
		    | \Psi_{\mathrm{gs}} \rangle,
		    E^N,
		    p, p', j, \sigma',
		    S_{p j, p j},
            N_{p' \sigma', p' \sigma'}, 
		    N_{\mathrm{meas}}
		    $}
		    \State $T^{\pm}_{p j, p' \sigma'} :=$
		        \textsc{AmplsSpinChargeAux}$(
		        \mathcal{H},
		        | \Psi_{\mathrm{gs}} \rangle, E^N,
		        \pm,
		        p, p', j, \sigma',
		        N_{\mathrm{meas}})$
    		\For{$\lambda$}
                \State $
                    M_{\lambda p j, p' \sigma'}
                    :=
                    e^{-i \pi/4}
                    T_{\lambda p j, p' \sigma'}^+
                    +
                    e^{i \pi/4}
                    T_{\lambda p j, p' \sigma'}^-
                    -
                    \sqrt{2}
                    S_{\lambda p j, p j}
                    -
                    \frac{N_{\lambda p' \sigma', p' \sigma'}}{2 \sqrt{2}}
                    $
                \Comment{See eq. (\ref{transition_mat_spin_charge_from_T})}
            \EndFor
            \State \Return $M_{p j, p' \sigma'}$
        \EndFunction
	\end{algorithmic}
\end{algorithm}

\begin{algorithm}[H]
	\caption{Calculation of transition amplitudes for auxiliary states}
    \label{alg:AmplsSpinChargeAux}
	\begin{algorithmic}[1]
		\Function{AmplsSpinChargeAux}{$
		    \mathcal{H},
		    | \Psi_{\mathrm{gs}} \rangle,
		    E^N,
		    \nu,
		    p, p', j, \sigma',
		    N_{\mathrm{meas}}
		    $}
            \State $T^{\nu}_{p j, p' \sigma'} := 0$
    		\For{$m = 1, \dots, N_{\mathrm{meas}}$}
    		    \State Input $| \Psi_{\mathrm{gs}} \rangle$ to $\mathcal{C}_{p j, p' \sigma'}^\nu$ and measure the ancillae
    		        $| q_2^{\mathrm{A}} \rangle$ and
    		        $| q_0^{\mathrm{A}} \rangle$
    		    \Comment{Circuit in Fig. \ref{circuit_prep_el_xy_ab_aux_using_U}}
                \If{$
                    | q_2^{\mathrm{A}} \rangle
                    \otimes
                    | q_0^{\mathrm{A}} \rangle
                    ==
                    | 0 \rangle
                    \otimes
                    | 0 \rangle
                    $}
        		    \State $E :=$ QPE$(| \widetilde{\Psi} \rangle, \mathcal{H})$,
            		    Find $E$ among $\{ E_\lambda^N \}_\lambda$
        		    \Comment{For the register $| \widetilde{\Psi} \rangle$ coming out of the circuit}
    		        \State $\switch T^\nu_{\lambda p j, p' \sigma'} += 1$
    		    \EndIf
    		\EndFor
            \State $\switch
                T^{\nu}_{p j, p' \sigma'} *= 1/N_{\mathrm{meas}}$
            \State \Return $T^{\nu}_{p j, p' \sigma'}$
        \EndFunction
	\end{algorithmic}
\end{algorithm}

\end{widetext}

\bibliographystyle{apsrev4-1}
\bibliography{ref}

\end{document}